\documentstyle[12pt,aaspp4,flushrt]{article}

\def\K{\hskip 2pt K\hskip 3pt}
\def\nhi{N_{\rm HI}}
\def\cm{{\rm\,cm}}
\def\km{{\rm\,km}}

\def\kpc{{\rm\,kpc}}
\def\mpc{{\rm\,Mpc}}
\def\sec{{\rm\,s}}

\def\kms{\ifmmode{\,{\rm km\,s}^{-1}} \else {${\rm\,km\,s}^{-1}$}\fi}

\def\lya{{\rm Ly}\alpha}

\def\hmpc{\ifmmode{h\,{\rm Mpc}^{-1}} \else $h\,{\rm Mpc}^{-1}$\fi}
\def\vv606{V_{606}}
\def\i814{I_{814}}
\def\wtheta{{\omega(\theta)}}

\begin{document}

\title{\bf Annual Report \\
July 1, 1997 -- June 30, 1998}

\author{Princeton University Observatory\\
Peyton Hall\\
Princeton, NJ  08544 USA}

\date{\today}

\section{PERSONNEL}

\noindent
During the reporting year the research staff consisted of Neta Bahcall, Renyue
Cen, Bruce Draine, Edward Fitzpatrick, Jeremy Goodman, J. Richard Gott, James
Gunn, Edward Jenkins, Gillian Knapp, Russell Kulsrud, Jeremiah Ostriker,
Bohdan Paczy\'nski, David Spergel, Michael Strauss, Scott Tremaine and Edwin
Turner, as well as postdoctoral fellows Greg Bryan, Alex Lazarian, \v{Z}eljko
Ivezi\' c, Todd Tripp, Michael Vogeley, and Yun Wang. Christophe Alard
(Cambridge) and Janusz Kaluzny (Warsaw University Observatory) held visiting
research appointments during summer and fall 1997. Edward Fitzpatrick left in
summer 1997 to take up a faculty position at Villanova University, and Michael
Richmond left in summer 1997 to begin a faculty position at the Rochester
Institute of Technology. Tatsushi Suginohara returned to the University of
Tokyo in March 1998. Turner was on research leave during the fall term, and
Goodman and Paczy\'nski were on leave during the spring.

Neta Bahcall was inducted into the National Academy of Sciences.  J. Richard
Gott received Princeton's President's Award for Distinguished Teaching, as
well as the highest award of the Astronomical League, the national association
of amateur astronomers. Scott Tremaine received the Dannie Heinemann Prize for
Astrophysics from the American Astronomical Society and the Dirk Brouwer Award
from the Division of Dynamical Astronomy of the AAS. Two special celebrations
were held: the department hosted a successful festschrift celebrating Russell
Kulsrud's 70th birthday, and the Lyman Spitzer Building was dedicated at the
Princeton Plasma Physics Laboratory.

The 1998 Lyman Spitzer, Jr. Lecture Series was given by Reinhard Genzel on
``Galaxies and galactic nuclei in the infrared''. 

\section{RESEARCH PROGRAM}

\subsection{Cosmological models}

\noindent
N. Bahcall and X. Fan (graduate student) summarized the constraints on the
mass density of the universe as determined from several independent
observational methods.  The methods include the observed masses and
mass-to-light ratios of galaxies, groups and clusters as a function of scale;
the observed baryon fraction in clusters; and the evolution of cluster
abundance.  Their summary, ``A Lightweight Universe?" was published as an
Inaugural Article in the Proceedings of the National Academy of Sciences.
They showed that these independent methods all indicate that the mass density
of the universe is sub-critical, insufficient to halt the universal expansion,
and reveal a consistent picture of a lightweight universe with only
$\sim$20--30\% of the critical density.  Thus, the universe is likely to
expand forever.

D. Spergel, N. Cornish (Cambridge), and G. Starkman (Case Western) are
exploring the possibility that the universe has non-trivial
topology. Observations of fluctuations in the microwave cosmic background
radiation (CBR) can yield information not only about the geometry of the
universe, but also about its topology. If the universe has negative curvature,
then the characteristic scale for the topology of the universe is the
curvature radius. Thus, if we are seeing the effects of the geometry of the
universe, we can hope to soon see signatures of its topology. The cleanest
signature is written on the microwave sky: there should be thousands of pairs
of matched circles. These circles could be used to determine the precise
topology and volume of the universe.

W. Colley (graduate student) has measured the topology of the large-scale
structure observed in the Las Campanas Redshift Survey (LCRS).  This survey
provides deep slices showing galaxies out to a redshift $z \simeq 0.2$, giving
us our best current picture of the structure of the universe on the largest
scales.  One can see many voids and walls. Colley filtered these data with a
smoothing length of $20 h^{-1}\mpc$ ($h=H_0/100\kms\mpc^{-1}$, where $H_0$ is
the Hubble constant) so as to recover fluctuations that are still in the
linear regime.  Thus, if the original fluctuations were Gaussian random-phase,
so should be the topology of the smoothed maps.  He compared the topology as
measured by the genus, the number of high-density spots minus the number of
low-density spots, as a function of density threshold.  The results are in
extraordinarily good agreement with the theoretical random-phase curve,
showing that this pattern of galaxy clustering could have grown by
gravitational instability from random quantum fluctuations in a standard
inflationary Big-Bang model.

J. R. Gott and L.-X. Li (graduate student) have written two papers considering
quantum effects in spacetimes with multiply connected topologies and closed
timelike curves.  A previous calculation of the renormalized energy-momentum
tensor in Misner space showed a blowup as the Cauchy horizon is approached,
which was the basis for Hawking's Chronology Protection Conjecture forbidding
time travel. However, this model has a non-self-consistent vacuum---an adapted
Minkowski vacuum, which does not solve Einstein's field equations for the
Misner geometry. In their first paper, Li and Gott showed that if, instead, an
adapted Rindler vacuum is used one can obtain a self-consistent vacuum where
the renormalized energy-momentum tensor is well behaved as the Cauchy horizon
is approached and self-consistently solves Einstein's field equations for
Misner space.  The second paper explores whether the laws of physics permit
the universe to be its own mother.  An inflationary universe can create baby
universes. If one of those baby universes turns out to be the inflationary
universe that one started with, then there is an early region of closed
timelike curves in the model.  In such a case the universe has no earliest
event.  Gott and Li show how there are self-consistent vacuum states, and how
an arrow of time with pure retarded potentials naturally arises in such
models.

M-G. Park and Gott have shown how a study of gravitational lens
separations as a function of quasar redshift can yield data on the curvature
of the universe. The observed separations become smaller with increasing
source redshift, and Park and Gott show that the observations are not
compatible with the ``standard'' gravitational lensing statistics model in a
flat universe. They also tried various open and flat cosmologies, galaxy mass
profiles, galaxy merging and evolution models, and lensing aided by clusters
to explain the correlation, but the data are not compatible with any of
these possibilities within the 95\% confidence limit.
   
\subsection{Cosmic background radiation}

\noindent
D. Spergel is a member of the science team for the Microwave Anisotropy Probe
(MAP) satellite (http://map.gsfc.nasa.gov).  Fluctuations in the CBR trace the
density fluctuations 300,000 years after the Big Bang. The COBE satellite has
already established that these fluctuations are present at a level consistent
with the gravitational growth of structure from small primordial
fluctuations. MAP will survey the CBR temperature and polarization at 5
frequencies ranging from 22--90 GHz, at angular resolution as high as
$0.21^\circ$. These high-resolution maps should determine whether the
primordial fluctuations are adiabatic density variations (as predicted by
inflation) or isocurvature fluctuations (as predicted in alternative models of
structure formation). If the fluctuations are adiabatic, MAP should yield
precision measurements of the geometry of the universe, the ratio of the
density in baryons to the density in non-baryonic dark matter, the ratio of
dark matter density to radiation density, as well as the primordial spectrum
of fluctuations. If the fluctuations are not adiabatic, then we will have to
rethink our standard ideas about the origin of structure.

S. P. Oh (graduate student), Spergel and G. Hinshaw (Goddard) developed an
efficient technique to determine the power spectrum from CBR sky maps.
Existing algorithms for computing the angular power spectrum of a pixelized
map typically require O($N^3$) operations and O($N^2$) storage, where $N$ is
the number of independent pixels. The MAP and Planck satellites will produce
megapixel maps of the CBR temperature at multiple frequencies; thus existing
algorithms are not computationally feasible. The new algorithm requires only
O($N^2$) operations and O($N^{1.5}$) storage, representing a million-fold
speedup in determining the power spectrum from MAP or Planck.

A. Refregier, Spergel and T. Herbig (Physics) analyzed the extragalactic
foregrounds for the MAP mission.  While the major contributor to CBR
anisotropies is the sought-after primordial fluctuations produced at the
surface of last scattering, other effects can also be important.  They show
that the dominant ``noise'' sources are the Sunyaev-Zel'dovich effect from
rich clusters and gravitational lensing. MAP should detect about 50 discrete
sources and 10 clusters directly. Fainter clusters can be probed by
cross-correlating MAP with cluster positions extracted from existing catalogs.
Finally, they consider probing the hot gas on supercluster scales by
cross-correlating the CBR with galaxy catalogs. Assuming that galaxies trace
the gas, they show that a cross-correlation between MAP and the APM galaxy
catalog should yield a marginal detection, or at least a four-fold improvement
on the COBE upper limits for the rms Compton y-parameter.

Y. Wang, Spergel and E. Turner have investigated the implications of
anisotropies in the CBR for possible large-scale variations in Hubble's
constant.  Low-amplitude (linear regime) cosmic density fluctuations lead to
variations in the locally measurable value of $H_0$ (denoted as $H_L$). For
three currently viable structure formation models based on cold dark matter
(CDM)---tilted CDM, $\Lambda$CDM, and open CDM---normalized by the 4-year COBE
DMR data, the fractional variations in $H_L$ are of order $3$--$6$\% (95\%
confidence interval) in a sphere $200h^{-1}\mpc$ in diameter, and of order
$1$--$2$\% if the size of this sphere is doubled.  The measured CBR dipole
(caused by the Galaxy's peculiar velocity with respect to the CBR rest frame)
provides additional constraints on $\Delta H$ that supplement our limited
knowledge of the power spectrum. In the limiting case where the power spectrum
is unknown, the observed CBR dipole alone provides a very robust limit,
$\langle \delta_H^2 \rangle_R^{1/2} <10.5\, h^{-1}\mpc/R$ in a sphere of
radius $R$ (95\% confidence level).  Thus, variations between currently
available local measures of $H_0$ and its true global value of a few to
several percent are to be expected and differences as large as 10\% are
possible.

\subsection{Large-scale structure}

\noindent
J. Ostriker summarized the development of large-scale cosmic structure at the
1998 San Diego meeting of the American Astronomical Society.  There is now a 
standard model for the growth of structure in the universe which is
clearly correct in general form, although the details remain
uncertain. Quantum fluctuations in the early universe, amplified by inflation,
are probably the original source of perturbations in an otherwise uniform
universe. The amplitude of these perturbations is determined by observed CBR
fluctuations and the shape is predicted by standard theory, so long as some
form of CDM constitutes most of the mass density.  Putting these
initial conditions, with standard atomic physics and the equations of
hydrodynamics, into a sufficiently detailed computational model, allows one to
predict where and when galaxies will form. Very large scale simulations
completed to date of models such as the concordance (Ostriker and Steinhardt)
model indicate quite good agreement with a suite of observational constraints
from $\lya$ clouds at redshift 3 to clusters of galaxies at redshift
zero. Open models with $\Omega_m<1$ seem best at reproducing all of these
properties, as well as the large-scale distribution of galaxies and
intergalactic gas. A particularly interesting and robust conclusion is reached
concerning the distribution of baryons. For a wide range of viable
cosmological models, the fraction of baryons that condenses to stars and cold
gas (in galaxies) is $20\pm 10\%$. Of the remaining baryons in the
intergalactic medium, the fractions in the hot ($T> 10^7\K$), intermediate
($10^7\K >T>10^5\K$) and warm ($10^5K>T$) components are $20\pm7\%$,
$55\pm10\%$, and $15\pm5\%$ respectively.  Thus the majority of cosmic baryons
are found in intermediate-temperature regions ($T\sim10^6\K$), where they
should be discovered by soft X-ray observations.

R. Cen has examined all currently viable CDM models in great detail to find
ways to differentiate between them. The COBE measurements of CBR fluctuations
and the local abundance of rich clusters of galaxies provide the two most
powerful constraints on cosmological models.  When all variants of the
standard CDM (SCDM) model are subjected to the combined constraints, the power
spectrum of any model is fixed to $\sim 10\%$ accuracy in both shape and
overall amplitude.  These constrained models are not expected to differ
dramatically in their local large-scale structure properties.  However, their
evolutionary histories differ with the differences growing dramatically with
redshift.  The observational constraints include the correlation function of
rich clusters of galaxies, the galaxy power spectrum, the evolution of cluster
abundance, gravitational lensing by moderate-to-high redshift clusters, the
$\lya$ forest, damped $\lya$ systems, high-redshift galaxies, reionization of
the universe and future CBR experiments.  The combined power of several or all
of these observations is tremendous.  Thus, we appear to be on the verge of
being able to make dramatic tests of all models in the near future using a
rapidly growing set of observations, mostly at moderate to high redshift.

Ostriker, M. Blanton (graduate student), Cen and M. Strauss used large-scale
hydrodynamic simulations (with heuristic criteria for galaxy formation) to
investigate the bias factor $b(R)=\sigma_g(R)/\sigma(R)$, where $\sigma_g(R)$
is the variance of galaxy counts in spheres of radius $R$ and $\sigma(R)$ is
the same for mass. They found that the bias factor varies from 2.6 at 1
$h^{-1}\mpc$ to 1.2 at $30 h^{-1}\mpc$.  Including the dependence of the
galaxy density on local gas temperature as well as on local mass density can
fully account for this scale dependence.  Galaxy density depends on
temperature because gas which is too hot cannot cool to form galaxies; this
causes scale dependence of $b(R)$ because local gas temperature is related to
the gravitational potential, and thus contains information about the
large-scale density field.  They also found that the relationship between the
galaxy and mass density fields is a function of galaxy age.  On large scales,
older galaxies are highly biased ($b\sim1.7$) and strongly correlated ($r
\sim1.0$) with the mass density field; younger galaxies are not biased
($b\sim0.8$) and only weakly correlated ($r\sim 0.5$) with the mass.  Thus
linear bias is inadequate to describe the relationship between galaxies and
mass.

Wang, Spergel and Strauss have explored combining data from MAP with data from
the upcoming Sloan Digital Sky Survey (SDSS) to constrain inflationary
models. The existence of primordial adiabatic Gaussian random-phase density
fluctuations is a generic prediction of inflation. The properties of these
fluctuations are completely specified by their power spectrum. The basic
cosmological parameters and the primordial power spectrum together completely
specify predictions for the CBR fluctuations and large-scale structure.  If we
assume that the cosmological parameters are known a priori, the combined data
from MAP and SDSS can constrain the primordial power spectrum to $\sim 10\%$
accuracy for $k\sim 0.01h\mpc^{-1}$, and to $\sim 1$\% accuracy for $k\sim
0.1h\mpc^{-1}$. The uncertainty in the primordial power spectrum increases by
less than a factor of 2 if we solve simultaneously for the cosmological
parameters $h$, $\Lambda$, $\Omega_b, \tau_{ri}$, and the effective bias
$b_{\rm eff}$ between the matter density field and the galaxy redshift density
field.

L. A. Phillips (graduate student) and Turner have analyzed bright-end ($K =
10$--17) galaxy counts from a number of near-infrared galaxy surveys.  All
these surveys agree that the observed near-infrared galaxy number counts are
inconsistent with a simple no-evolution model.  They examined evolutionary
effects and a local underdensity as possible causes of this effect and
found that the data are fit by either a factor of 1.7--2.4 deficiency of
galaxies out to redshift $z = 0.10$--$0.23$, or by unexpectedly strong
low-redshift evolution in the K-band, leading to corrections at $z= 0.5$
that are as much as 60\% larger than accepted values.  The former possibility
would imply that the local expansion rate on scales of several hundred Mpc
exceeds the global value of $H_0$ by up to 30\% and that the amplitude of very
large-scale density fluctuations is far larger than expected in any current
cosmogonic scenario.  The latter possibility would mean that even the
apparently most secure aspects of our understanding of galaxy evolution and
spectral energy distributions are seriously flawed.

Wang, Bahcall and Turner derived empirical color-redshift relations for
galaxies in the Hubble Deep Field (HDF).  The dispersion between the estimated
redshifts and the spectroscopically observed ones is small, ranging from
$\sigma_z\simeq 0.03$ to 0.1 for $z < 2$ galaxies, and from $\sigma_z\simeq
0.14$ to 0.36 for $2 < z < 4$.  They applied the color-redshift relations to
the HDF photometric catalog and obtained estimated redshifts that are
consistent with those derived from spectral template fitting methods.  The
advantages of the color-redshift relations are that they are simple to use and
do not depend on the assumption of particular spectral templates; they provide
model-independent redshift estimates for $z < 4$ galaxies using only
multi-band photometry. They have used these results to investigate the
redshift distribution of galaxies in the HDF and found peaks that suggest
large-scale clustering of galaxies to at least $z\sim 1$, consistent with
those identified in spectroscopic probes of the HDF.

Strauss carried out a variety of observational and theoretical analyses of
large-scale structure.  In collaboration with L. Guzzo (Milan), K. Fisher
(Texas), R. Giovanelli and M. Haynes (Cornell), he analyzed the clustering
statistics and small-scale velocity dispersion of subsamples of galaxies
selected by morphology from the Pisces-Perseus redshift survey.  They showed
that the small-scale velocity dispersion is a strong function of morphological
type.  With R. Kim (graduate student), he developed a new method for measuring
skewness and kurtosis from redshift surveys, which is appreciably more robust
than the standard moments technique used in the literature. With Ostriker and
Cen, he developed a new measure of the small-scale velocity dispersion, making
it explicitly a function of local density.  They showed directly what had been
suspected for years: that the velocity field in the field (i.e., outside of
clusters) is very quiet.  This is a powerful new statistic to distinguish
cosmological models. With D. Goldberg (graduate student), he carried out an
analysis of the effect of baryon density on the average-scale power spectrum of
density, and showed that (with luck) it should be a measureable effect on the
SDSS power spectrum.  With Y. Sigad, A. Eldar, A. Dekel (Hebrew University)
and A.  Yahil (Stony Brook), he made a detailed comparison of the observed
density distribution of IRAS galaxies with that inferred from 
peculiar velocities.  The comparison is consistent with gravitational
instability theory, allowing a constraint on the quantity $\beta \equiv
\Omega^{0.6}/b= 0.89 \pm 0.12$.

Vogeley showed that statistical analysis of the unresolved light in the Hubble
Deep Field (HDF) strongly constrains possible sources of the optical
Extragalactic Background Light (EBL).  This constraint is crucial for
determining the spectrum of the EBL because reported upper limits on the
optical EBL are several times larger than the surface brightness from detected
galaxies, suggesting the possibility of additional galaxy populations.  To
test for the statistical signature of previously undetected sources, he
estimated the auto-, cross-, and color correlations of the ``sky'' in the HDF
that remains after masking objects brighter than $\i814=30$ mag.  Auto- and
cross-correlations of surface brightness in the $\vv606$ and $\i814$ bandpasses
are well-fitted by $\wtheta \sim 10^{-6}(\theta/1'')^{-0.6}$ up to $10''$.
This measurement yields the most stringent limits to date on small-scale
structure in the night sky.

An important implication of Vogeley's work is that, unless there is a truly
uniform optical background, the mean EBL is likely to be within a small
fraction of the surface brightness from detected galaxies.  No currently
plausible sources of additional EBL satisfy the constraints that they (1)
would not have already been detected, (2) contribute EBL comparable to that
from detected galaxies, and (3) do not produce EBL fluctuations in excess of
the upper limits set by correlations in the HDF.  These constraints admit only
a confusion-limited population of extremely low surface-brightness objects
that is disjoint from the parameter space of all detected galaxies.
Extrapolation of detected galaxy counts to zero flux would add only a few
percent to the EBL.  Diffuse intergalactic light clustered similarly to faint
galaxies could explain some of the observed correlations but would contribute
at most a few $\times 10\%$ to the mean EBL.

Vogeley initiated a project with J. Gunn to further constrain fluctuations in
the EBL by combining analysis of deep Hubble Space Telescope (HST) WFPC2
imaging with ground-based imaging to be taken in drift-scan mode with the
Apache Point Observatory 3.5-meter telescope.  This analysis will be applied
to the HDF and other deep HST images, including the HDF South.

Vogeley reviewed measurements of the power spectrum of density fluctuations
from galaxy redshift surveys and discussed advances that will be possible with
the SDSS.  This review emphasized the difficulties of high-precision power
spectrum estimation in the presence of Galactic extinction, photometric
errors, galaxy evolution, clustering evolution, and uncertainty about the
background cosmology. Discussed in this review are some of the ways in which
the SDSS seeks to overcome these obstacles.

Strauss and Vogeley, with M. Tegmark (IAS), A. Hamilton (Colorado), and
A. Szalay (Johns Hopkins) showed how precision measurements of the galaxy
power spectrum $P(k)$ require a data analysis pipeline that is both fast
enough to be computationally feasible and accurate enough to take full
advantage of high-quality data.  To improve speed, Karhunen-Lo\`eve
power-spectrum estimation can be accelerated with a quadratic data compression
scheme.  To improve accuracy, they derived analytic expressions for handling
the integral constraint, since it is crucial that finite volume effects are
accurately corrected for on scales comparable to the depth of the survey.
Also shown is that for data analysis methods based on counts in cells, such as
the Karhunen-Lo\`eve and quadratic techniques, multiple constraints can be
included via simple matrix operations, thereby rendering the results less
sensitive to galactic extinction and mis-estimates of the radial selection
function.  They describe a data analysis pipeline that does justice to the
increases in both quality and quantity of data that upcoming redshift surveys
will provide.  It involves using three analysis techniques in conjunction: a
traditional Fourier approach on small scales, a pixelized quadratic matrix
method on large scales and a pixelized Karhunen-Lo\`eve eigenmode analysis to
probe anisotropic effects such as redshift space distortions, residual
extinction, and radial and selection function errors.

Vogeley and A. Connolly (Johns Hopkins) analyzed the effect of uncertainty in
corrections for Galactic extinction and zero-point extinction and photometry
errors on estimates of the three-dimensional power spectrum from the SDSS and
other deep surveys.  Galactic extinction or photometric zero-point
fluctuations cause erroneous fluctuations in the redshift-space distribution
of galaxies, hence extra apparent clustering power on scales of order the
angular size of these variations at the effective depth of the survey. For the
SDSS, errors in the extinction map used to correct apparent galaxy magnitudes
can be the dominant source of systematic error on wavelength scales of
$\lambda >300h^{-1}\mpc$, just where it is hoped that the SDSS will provide a
crucial link between our current knowledge of fluctuations in the
present-epoch galaxy distribution and the mass distribution at redshift
$z=10^3$ probed by CBR anisotropy experiments. These analyses provide
guidelines for the accuracy with which the SDSS and other similar surveys must
correct for extinction and for the large-scale photometric calibration.

Cen, in collaboration with J. Einasto (Estonia), M. Einasto (Estonia), E. Tago
(Estonia), A. A. Starobinsky (Moscow), F. Atrio-Barandela (Spain), V. M\"uller
(Potsdam), A. Knebe (Potsdam), P. Frisch (G\"ottingen), H. Andernach (Mexico)
and D. Tucker (Fermilab), has analyzed the power spectrum of galaxies using
published data. On intermediate and small scales the power spectrum is best
given by the two-dimensional distribution of APM galaxies. This sample is not
influenced by redshift distortions and is the largest and deepest sample of
galaxies available.  On large scales they use power spectra derived from
three-dimensional data, which are reduced in amplitude to the power spectrum
of APM galaxies.  They find that the available data indicate the presence of
two different populations in the local universe. Samples of clusters of
galaxies as well as the APM 3-D, IRAS QDOT, and SSRS+CfA2 galaxy surveys cover
relatively large regions in the universe where rich, medium and poor
superclusters are well represented. The mean power spectrum of these samples
has a relatively sharp maximum at wavenumber $k=0.05 \pm 0.01\hmpc$, followed
by a power-law spectrum of index $\approx -1.9$ toward smaller scales.  The
power spectrum found from LCRS data represents regions of the universe with
medium-rich and poor superclusters; it is flatter around the maximum. They
argue that the former power spectrum probably corresponds to a fair sample of
the universe.

Cen, in collaboration with J. Einasto (Estonia), M. Einasto (Estonia), E. Tago
(Estonia), V. M\"uller (Potsdam), A. Knebe (Potsdam), F. Atrio-Barandela
(Spain) and D. Tucker (Fermilab), suggests a new method to find $\sigma_8$,
the rms mass fluctuation in a sphere of radius $8h^{-1}\mpc$. The method is
based on an integration of the mean power spectrum of galaxies, which is then
reduced to a power spectrum of the mass using a simple relation based on the
fraction of mass in galaxies, which is determined from detailed modeling of
void evacuation for various cosmological models. They find $\sigma_8=0.89 \pm
0.05$ for galaxies, and $\sigma_8 = 0.68 \pm 0.06$ for mass.

\subsection{Intergalactic medium and galaxy formation}

\noindent
G. Bryan, in collaboration with M. Machacek (Northeastern), P. Anninos and
M. Norman (NCSA), used Eulerian hydrodynamic simulations to study the
properties of the $\lya$ forest at redshifts $z\simeq 2$--5.  As a first step
towards using these simulations to constrain cosmological parameters, they
performed a numerical resolution study to examine the robustness of the
predicted quantities.  They found that the column density distribution is
relatively insensitive to spatial resolution, but that the distribution of the
Doppler b-parameter, which is a measure of linewidth, depends sensitively on
resolution, decreasing as the resolution increases.  This is important because
the new predicted distribution for the SCDM model is now significantly lower
than observed.  The same group, along with A. Meiksin (Edinburgh), has begun
the second step of this project, to use simulations of a number of
cosmological models to constrain their parameters.  In particular, they have
found that the shape of the column density distribution is a sensitive probe
of the amplitude of fluctuations on the $\sim 0.5\mpc$ scale at $z \sim 3$.
The observed shape appears to favor models with more small-scale power (such
as models including massive neutrinos or a tilted initial spectrum), over
those that have less power on small scales.  They have also found that the
distribution of b-parameters is sensitive to the power spectrum.

J. Kepner (graduate student), Bryan, and Spergel have examined the
formation of galaxy-sized objects in hydrodynamic simulations, using a new
adaptive-mesh method that combines the shock-capturing ability of Eulerian
hydrodynamics codes with adaptive mesh placement for high resolution in
regions of large density.  They examined the formation and evolution of
cold-gas disks (star formation was not included) and found that, as previously
predicted, they formed inside-out by accreting high-angular momentum material
at late times.  Two cosmological models were used, SCDM and $\Lambda$CDM, and
the disk sizes and specific angular momenta for both were similar to those
observed.

Bryan collaborated with T. Abel and Norman to explore the formation of the
first non-linear baryonic objects in models of hierarchical structure
formation.  These objects, with masses around $10^6 M_\odot$, form at $z \sim
20$ in most models.  They used an adaptive-mesh hydrodynamics code, and a
nine-species non-equilibrium chemistry model which included hydrogen, helium
and the species relevant for the formation of molecular hydrogen.  The clouds
were initially Jeans stable, but collapse occurred when the molecular hydrogen
fraction (which is the primary coolant since the gas composition is
primordial) reached $\sim 10^{-3}$.  There is evidence for fragmentation on
the scale of a few hundred solar masses.  Densities as high as $10^5\cm^{-3}$
and proper length scales as small as 0.02 pc (a spatial dynamic range of
$10^5$) were resolved.

Cen and R. Simcoe (undergraduate student) have analyzed the sizes, shapes and
correlations of $\lya$ clouds produced by a hydrodynamic simulation of a
spatially flat CDM universe with a non-zero cosmological constant
($\Omega_0=0.4$, $\Lambda_0=0.6$, $\sigma_8=0.79$), over the redshift range
$2\le z \le 4$.  The $\lya$ clouds range in size from several kpc to about a
hundred kpc, and in shape from round, high column-density regions with
$\nhi\ge 10^{15}$~cm$^{-2}$ to low column-density sheet-like structures with
$\nhi \le 10^{13}$~cm$^{-2}$.  The most common shape resembles a flattened
cigar.  The physical size of a typical cloud grows with time roughly as
$(1+z)^{-3/2}$ while its shape hardly evolves. These results demonstrate that
any simple model with a population of spheres (or other shapes) of a uniform
size is oversimplified; they also illustrate why the use of double quasar
sightlines to set lower limits on cloud sizes is useful only when the
perpendicular sightline separation is small ($\Delta r \le 50h^{-1}\kpc$).
Finally, they conjecture that high column-density $\lya$ clouds ($\nhi\ge
10^{15}\cm^{-2}$) may be the progenitors of the lower redshift faint blue
galaxies, since their correlation length, number density (extrapolated to
lower redshift) and masses are in fair agreement with those observed.

Cen, in collaboration with M. Rauch (Caltech), J. Miralda-Escud\'e (Penn),
W.L.W. Sargent (Caltech), T. A. Barlow (Caltech), D. Weinberg (Ohio State),
L. Hernquist (Santa Cruz), N. Katz (Massachusetts) and Ostriker, has measured
the distribution of the flux decrement caused by $\lya$ forest absorption from
intervening gas in the lines of sight to high-redshift quasars from a sample
of seven high-resolution quasar spectra obtained with the Keck Telescope. The
observed flux decrement distribution function (FDDF) is compared to the FDDF
from two simulations of the $\lya$ forest: a CDM model with $\Omega=0.4$,
$\Lambda=0.6$ ($\Lambda$CDM), computed with the Eulerian code of Cen and
Ostriker, and a standard CDM model with $\Omega=1$ (SCDM) computed with the
SPH code of Hernquist, Katz, and Weinberg.  Good agreement is obtained between
the shapes of the simulated and observed FDDFs for both simulations after
fitting only one free parameter, which controls the mean flux decrement.  The
difference between the predicted FDDFs from the two simulations is small, and
arises mostly from a different temperature in the low-density
gas (caused by different assumptions about the reionization history in the two
simulations), rather than differences between the two cosmological models {\it
per se}, or numerical effects in the two quite different codes.  A measurement
of the parameter $\mu \propto \Omega^2_{b}h^{3} / \Gamma$ (where $\Gamma$ is
the HI ionization rate due to the ionizing background) is obtained by
requiring the mean flux decrement in the simulations to agree with
observations.  Using a lower limit $\Gamma> 7\times10^{-13}\sec^{-1}$ from
the abundance of known QSOs, they derive a lower limit on the baryonic matter
density, $\Omega_{b} h^{2}>0.021$ ($0.017$) for the $\Lambda$CDM (SCDM)
model.  The measurement of $\mu (z)$ allows a determination of the evolution
of the ionizing radiation field with redshift; the models predict an intensity
that is approximately constant with redshift, which is in agreement with the
assumption that the ionizing background is produced by known quasars for $z <
3$, but requires additional sources of ionizing photons at higher redshift
given the observed rapid decline of the quasar abundance.

Cen, with S. Phelps (graduate student), Miralda-Escud\'e and
Ostriker, has examined the correlation function of $\lya$ clouds along the
line of sight in a $\Lambda$CDM cosmological model. A consistent picture seems
to emerge: the correlation strength for a given set of objects is positively
correlated with their characteristic global density, and the differences among
the correlations of galaxies, $\lya$ clouds and mass reflect the differences
in density that each trace.  They find that the galaxies are biased
over mass by a factor of $\sim 3.0$, in accord with recent observations of
high-redshift galaxies.  The correlation strength of $\lya$ clouds with column
densities of $10^{13}$--$10^{14}\cm^{-2}$ is comparable to that of the total
mass.  

Cen and Ostriker use high-resolution cosmological simulations of a
$\Lambda$CDM model to predict the distribution of
baryons at the present and at moderate redshift.  It is found that the average
temperature of baryons is an increasing function of time, with most of the
baryons presently having a temperature in the range $10^{5-7}$K.  Thus, not
only is the universe dominated by dark matter, but more than half of the
normal matter is yet to be detected.  Detection of this warm/hot gas poses an
observational challenge, requiring sensitive EUV and X-ray satellites.
Signatures include a soft cosmic X-ray background, apparent warm components in
hot clusters due to both intrinsic, warm, intra-cluster gas and inter-cluster
gas along the line of sight, absorption lines in X-ray and UV quasar spectra
[e.g., O VI (1032,1038)A lines, OVII 574eV line] strong emission lines (e.g.,
O VIII 653 eV line), and low-redshift, broad, low column-density $\lya$
absorption lines.

Tripp completed a study with L. Lu (Caltech) and B. Savage (Wisconsin) of the
relationship between low-redshift $\lya$ clouds and galaxies based on
HST spectroscopy of two QSOs, H1821+643 and PG1116+215.  The high
signal-to-noise of the HST spectra of these QSOs enables the detection of very
weak $\lya$ absorption lines: 26 $\lya$ lines with rest equivalent
width $W_{\rm r} >$ 50 m\AA\ are detected toward H1821+643 and 13 are detected
toward PG1116+215, which implies a density of 102$\pm$16 lines per unit
redshift. The two-point velocity correlation function of the $\lya$ clouds
shows marginal evidence of clustering on $\sim500\kms$ scales, but
only if the weakest lines are excluded.  Tripp et al. also used the WIYN
telescope to measure the redshifts of galaxies in the $\sim 1^{\circ}$ fields
centered on each QSO.  They find 17 galaxy-absorber pairs within projected
distances of 1 Mpc with velocity separations of $350\kms$ or less. Monte Carlo
simulations show that if the $\lya$ lines are randomly distributed, the
probability of observing this many close pairs is 3.6$\times 10^{-5}$. Other
statistical tests also indicate that the $\lya$ absorbers are not randomly
distributed with respect to the observed galaxies.  These observations suggest
that many low-$z$ $\lya$ clouds are due to gas in large-scale structures of
galaxies rather than the gaseous halos of individual galaxies.

Savage, Tripp, and Lu have also completed a study of the O VI absorption line
systems detected in the spectrum of H1821+643.  Two O VI absorbers are
observed toward this radio-quiet QSO, an intervening system and an
``associated'' system with $z_{\rm abs} \approx z_{\rm QSO}$.  Savage et
al. analyze the physical conditions in these absorbers.  They find that the
intervening system could be photoionized by the UV radiation from background
QSOs if the absorption arises in very low density diffuse gas with an extended
distribution (the diameter of the absorber must be greater than 300 kpc).
Alternatively, the O VI absorption lines could originate in collisionally
ionized hot gas.  Two galaxies are present at the redshift of the intervening
O VI absorber with projected distances of 100 and 350 kpc.  Therefore this O
VI absorption could be due to the hot intracluster medium of a group of
galaxies.

Tripp, Lu, and Savage have obtained a high signal-to-noise ground-based
spectrum of the high-redshift QSO HS1700+6416 to study the metal absorption
line systems.  This QSO will be observed by the Space Telescope Imaging
Spectrograph investigation team in 1998 in order to search for very highly
ionized gas traced by species such as Ne VIII and Mg X, and the ground-based
spectrum obtained by Tripp et al. will be crucial for the analysis of the HST
data.  Species such as Ne VIII are optimal for probing the distribution of hot
gas in the vicinity of galaxy clusters and groups.  This will provide an
important test of cosmological hydrodynamic simulations which predict the
presence of such hot gas regardless of the specific cosmology assumed.

Kepner, Tripp, T. Abel (Max-Planck-Institut), and Spergel have constructed
detailed models of gas in small dark matter halos in order to predict their
absorption line signatures for comparison to QSO absorption lines.  The
models include full radiative transfer and gas dynamics and show the effects
of a multiphase absorbing medium since the mini-halo develops a self-shielded
core as the UV background intensity decreases.  They find that the absorption
signatures of mini-halos are consistent with the observed absorption lines
in many cases.  However, this model cannot match the properties of the
intervening O VI absorber studied by Savage, Tripp, and Lu.

M. and T. Suginohara (Tokyo) and Spergel propose that we may be able to detect
$z > 10$ objects through their carbon, nitrogen and oxygen emission lines.  By
redshift of 10, star formation in the first objects should have produced
considerable amounts of carbon, nitrogen and oxygen. The submillimeter lines
of C, N and O redshift into the millimeter and centimeter bands (0.5 mm--1.2
cm), where they may be detectable. High spectral resolution observations could
potentially detect inhomogeneities in C, N and O emission, and see the first
objects forming at high redshift. They calculate the expected intensity
fluctuations and discuss the frequency and angular resolution required to
detect them.  At $1+z \sim 10$, the typical protogalaxy has a velocity
dispersion of $30\kms$ and angular size of 1 arcsecond. If CII is the
dominant coolant, then they estimate a characteristic line strength of $\sim
0.1\hbox{ K}\,\kms$.  If the intensity fluctuations are detected, they will
probe matter density inhomogeneity, chemical evolution and ionization history
at high redshifts.

\subsection{Clusters and groups of galaxies}

\noindent
Bahcall, in collaboration with Cen and Fan, continued her investigation of the
evolution of clusters of galaxies and the strong constraints that cluster
evolution places on cosmological models.  They showed that measuring the
redshift evolution of the number density of rich clusters breaks the classical
degeneracy between $\Omega$ (the mass density parameter of the universe) and
$\sigma_8$ (the amplitude of mass fluctuations on $8h^{-1}\mpc$ scales),
$\sigma_8\Omega^{0.5}\simeq0.5$, that follows from the present-day cluster
abundance.  Using the Press-Schechter approximation and cosmological
simulations, they found that low-$\sigma_8$ models evolve exponentially faster
than high-$\sigma_8$ models for a given cluster mass.  The strong dependence
on $\sigma_8$ arises because clusters represent rarer density peaks in
low-$\sigma_8$ models.  In contrast, the evolution rate at $z<1$ is relatively
insensitive to the density parameter $\Omega$ or to the exact shape of the
power spectrum.  Cluster evolution therefore provides a unique and powerful
method to determine $\sigma_8$.

Bahcall and Fan then used the three most massive clusters of galaxies observed
so far at $z>0.5$ to constrain $\Omega$ and $\sigma_8$.  They showed that the
existence of such massive clusters ($\sim$twice the mass of Coma) at these
early epochs---even the existence of the {\it single}\ most distant cluster at
$z=0.83$ (MS 1054-03), with its large gravitational lensing mass, high
temperature ($\sim$12 Kev), and large velocity dispersion---is sufficient to
establish powerful cosmological constraints.  They find that high-density,
$\Omega$=1 ($\sigma_8\simeq$0.5--0.6) Gaussian models are ruled out by these
data, since they predict $\sim10^{-5}$ massive clusters at $z>0.65$
($\sim10^{-3}$ at $z>0.5$) instead of the 1 (3) clusters observed.  They
find best fit values of $\Omega=0.2{+0.3\atop-0.1}$ and $\sigma_8=1.2\pm0.4$
(95\%).

Bryan, Norman and R. Sunyaev (Max-Planck-Institut) investigated the role of
turbulence in X-ray clusters.  They found that hierarchical models of
structure formation generically predict turbulent velocities of roughly
$400\kms$ in rich clusters.  They showed that this would be difficult to detect
through microwave or sub-mm observations of the kinematic Sunyaev-Zel'dovich
effect, but should be clearly seen as an additional source of broadening in
metal emission lines, which future missions, such as XMM, could detect.

Cen has used a large simulation of a realistic cosmological model (CDM with
$H_0=65\kms\mpc^{-1}$, $\Omega_0=0.4$, $\Lambda_0=0.6$, $\sigma_8=0.79$) to
study projection effects on observables of clusters of galaxies, including
richness, velocity dispersion, X-ray luminosity, total mass estimates (using
the virial theorem, X-rays and lensing), gas fraction and substructure. The
principal results include: (i) The average gas-to-mass ratio in clusters
appears to be 30-40\% higher than its actual value; moreover, the width of the
observed distribution of this ratio is entirely accounted for by projection
effects. The enhancement in gas-to-mass ratio due to projection is not quite
sufficient to reconcile the standard nucleosynthesis value of
$\Omega_{baryon}$ with $\Omega_0=1$ for any plausible value of $H_0$. (ii)
Rich cluster masses derived from X-ray temperatures or galaxy velocity
dispersions underestimate the true cluster mass by about 20\% on average, with
the former displaying a smaller scatter. Methods based on gravitational
lensing overestimate the true mass by only 5--10\% but have larger scatter
than X-ray masses. It appears that projection effects alone may account for
the discrepancies between various mass estimates for individual
clusters. (iii) Projection inflates substructure measurements in galaxy maps,
and most of the substructure observed in real clusters of galaxies may be due
to projection.  (iv) The X-ray luminosity of a cluster within a radius
$1.0h^{-1}\mpc$ is hardly altered by projection, rendering the cluster X-ray
luminosity function a very useful and simple diagnostic for comparing
observations with theoretical predictions. (v) The only meaningful way to
compare predictions of a cosmological model with cluster observations is to
subject clusters in a simulated universe to exactly the same observational
biases and uncertainties, and to compare the ``observed" simulated clusters
with real ones.

Cen has proposed the correlation function of rich clusters of galaxies as a
test of topological-defect cosmological models.  Textures initially are
randomly distributed on scales larger than their size, in sharp contrast to
the initial high-density peaks in Gaussian models which are already strongly
clustered before any gravitational evolution has occurred.  One thus expects
that the correlation of large cosmic objects such as clusters of galaxies in
the texture model should be significantly weaker than its Gaussian
counterpart. Cen shows that a texture model with $\Omega_0=1$ and bias $b=2$
(as required by cluster abundance observations) predicts a correlation length
$\le 6h^{-1}\mpc$ for clusters, independent of richness.  On the other hand,
the observed correlation length for rich clusters is $\ge 10h^{-1}\mpc$. It
thus appears that the global texture cosmological model or any random-seed
cosmological models are ruled out at a very high confidence.

Cen, in collaboration with F. Governato (Durham), B. Moore (Durham), J.
Stadel, G. Lake, and T. Quinn (Washington), has used the dynamics of the Local
Group and its environment as a unique test of cosmological models.  The
velocity field within $5h^{-1}\mpc$ of the Local Group is extremely ``cold'':
the deviation from a pure Hubble flow, characterized by the observed radial
peculiar velocity dispersion, is only $\sim60\kms$. They compare the local
velocity field with similarly defined regions extracted from N-body
simulations of universes dominated by CDM. They find that
neither $\Omega=1$ nor $\Omega=0.3$ CDM models can produce a single candidate
Local Group that is embedded in a region with such small peculiar velocities:
the typical dispersions are 300\---700\kms\ and 150\---300\kms\ 
respectively, several times the observed value.

Spergel, Refregier, and Vogeley initiated a project to study the effect of
galaxies on their environment and, in turn, the impact of this environment on
galaxy evolution, focusing on the most common structures in the universe:
groups of galaxies and the filamentary network of large-scale structure which
they trace.  The velocity dispersion of groups and filaments is comparable to
that of the galaxies themselves, thus interactions are frequent and the
intragroup medium may be indistinguishable from the filamentary gas.  The
appearance of a variety of phenomena on these scales, including morphological
segregation, variation of the dark matter to baryon ratio, and variation of
metallicity, indicate that study of these regions is critical to understanding
feedback between galaxies and their environment. The astrophysical origins of
galaxy ``biasing'' and the diffuse X-ray background lie in this regime.  Using
both analytic methods and N-body/hydrodynamical simulations, they are studying
energetics and dynamics in this setting and predicting observable
relationships between galaxy and gas properties.

\subsection{Gravitational lenses}

\noindent
Cen, Ostriker and J. Wambsganss (Potsdam) have developed a new ray-tracing
method to study gravitational lensing by three-dimensional mass distributions.
As an example, the method is applied to a standard CDM universe. The results
include estimates of the frequency of multiply imaged quasars, the
distribution of separations of multiple quasars, and the redshift distribution
of lenses; all as a function of quasar redshift. The ultimate goal is to apply
this method to a number of cosmogonic models and to eliminate models whose
gravitational lensing properties are inconsistent with observations.

B. Paczy\'nski and his associates worked mostly on a major observing program
aimed at detecting gravitational microlensing in the bulge of our Galaxy and
in the Magellanic Clouds (the Optical Gravitational Lensing Experiment or
OGLE) and related projects. All of the hardware and most of the software
needed for OGLE was developed at Warsaw University Observatory, under the
leadership of A. Udalski.  The system became fully operational in January
1997.  In the spring of 1998 a new ``Early Warning System'' was implemented. A
total of 25 microlensing event candidates in the galactic bulge have been
detected up to July 27. Considerable effort is devoted to bringing the OGLE
data into the public domain as soon as practical, making heavy use of the
Web. 

The OGLE team, with Princeton graduate student P. Wozniak, confirmed the
``short'' distance (distance modulus $18.2$) to the Large Magellanic Cloud
using RR Lyrae variables and red clump giants.
   
Paczy\'nski pointed out that the proposed Space Interferometry Mission (SIM)
will be capable of determining the masses of the unknown halo objects
responsible for gravitational microlensing events, and of measuring directly
the masses of nearby, high proper-motion stars.

\subsection{Gamma-ray bursts, active galactic nuclei, and black holes}

\noindent
The remarkable afterglow of the gamma-ray burst of 8 May 1997 (GRB 980508) was
detected at X-ray, optical, and radio wavelengths.  For the first time,
absorption lines were seen, thus proving the cosmological distance of the
afterglow and vindicating Paczy\'nski's claim that GRBs are at cosmological
distances.  Shortly after this discovery, J. Goodman pointed out that the
radio source might show evidence of diffractive scintillation and that if so,
this observation could provide an upper limit to the angular size of the
source and hence a lower limit to its distance.  Soon after Goodman's
prediction, Frail et al. announced radio flux variations that appear
consistent with scintillation, implying a source size of a few microarcseconds
and a relativistic expansion velocity at the time of the radio detection.

Paczy\'nski pointed out that the known afterglows of gamma-ray bursts indicate
that the bursts are located in or near star-forming regions.  Several new
positions point in the same direction, indicating that the bursts are related
to violent deaths of young massive stars, and not to the mergers of old
neutron stars.  Neutron star mergers are likely to result not in bursts but
rather in short optical transients.

Simple models developed by Li and Paczy\'nski indicate that neutron star
mergers are likely to result not in gamma-ray bursts but in optical transients
located far from the centers of their host galaxies.  Recently such optical
transients, with no apparent host galaxies, were discovered in deep supernova
searches.  These events may be related to neutron star mergers.

A. Ulmer (formerly a Princeton graduate student, now a postdoc at
Max Planck Institut f\"ur Astrophysik), Paczy\'nski, and Goodman have
re-examined optically thick atmospheres formed from stars tidally
disrupted by supermassive black holes.  A. Loeb (Harvard) and Ulmer
had concluded in 1997 that the effective temperature of such envelopes
has a roughly constant universal value depending only weakly on the
mass of the black hole.  In the new study, the photospheric radius and
effective temperature are shown to be extremely sensitive to an
ill-determined inner boundary condition near the black hole.
Nevertheless, the temperature has a minimum comparable to the
previously proposed universal value and favorable for optical
detection---about 6000 K. Remarkably, mild general-relativistic effects
cause convection in such atmospheres. 

A. H. Diercks, E. W. Deutsch (Washington), F. J. Castander, D. Q. Lamb
(Chicago), C. Corson (APO), G. Gilmore (Cambridge), R. Wyse (Johns Hopkins)
and Turner exploited the remote observing and fast instrument change
capabilities of the Apache Point Observatory (APO) 3.5-meter telescope to
obtain R-band and J-band light curves of an optical transient which is likely
to be associated with the gamma-ray burst event GRB 971214.  Their first
measurement took place only 17 hours after the gamma-ray event. The brightness
decayed with a power-law exponent of approximately $-1.2$, similar to but
slightly steeper than those of two previous well observed events (GRB 970228
and GRB 970508). The transient decayed monotonically during the first four
days following the gamma-ray event---this contrasts with the optical transient
associated with GRB 970508, which peaked in brightness two days after the
burst.

Turner has collaborated with A. Yonehara, S. Mineshige (Kyoto), J. Fukue
(Osaka Kyoiku) and M. Umemura (Tsukuba) to study microlensing diagnostics of
accretion disk structure.  The optical/ultraviolet continuum from active
galactic nuclei (AGN) seems to originate from optically thick and/or thin
disks, and occasionally from associated circumnuclear starburst regions.
These different possible origins can, in principle, be distinguished by
observations of gravitational microlensing events.  They performed numerical
simulations of microlensing of an AGN disk by a single star passing in front
of the AGN. The calculated spectral variations and light curves show very
different behavior for the different models: variability over a few months
with strong wavelength dependence is expected in the case of an optically
thick disk (standard model), while an optically thin disk (advection-dominated
model) will produce shorter, nearly wavelength-independent variation. With a
starburst region, much slower variations (over a year) are superimposed on
the shorter variations.

Turner continued to work with A. Yonehara, S. Mineshige, T. Manmoto (Kyoto),
J. Fukue (Osaka Kyoiku), and M. Umemura (Tsukuba) to develop an X-ray
microlensing probe of accretion disk structure in the gravitationally lensed
quasar Q2237+0305.  The innermost regions of the quasar can be resolved by a
gravitational-lens ``telescope" on scales down to a few AU.  For this purpose,
X-ray observations are ideal, because X-rays originate from the innermost
regions of the disk.  They performed numerical simulations of microlensing of
a standard optically thick disk as well as an optically-thin,
advection-dominated accretion flow. They find that X-ray radiation, which is
produced in optically thin regions, exhibits intensity variation over a few
tens of days. In contrast, optical-UV fluxes, which are likely to come from
the optically thick region, exhibit more gradual light changes, which are
consistent with the microlensing events so far observed in Q2237+0305.
Currently, Q2237+0305 is being monitored for lensing events at APO. Once a
lensing event begins, simultaneous multi-wavelength observations by X-ray
satellites (e.g., ASCA, AXAF, XMM) as well as the Hubble Space Telescope (HST)
could reveal an AU-scale central accretion disk around a black hole.

Blanton, Turner and J. Wambsganss (Potsdam) analyzed
observations of the quadruply lensed quasar Q2237+0305 which they obtained
with HST.  On a timescale of 3--4 hours, they observed no variation in
component A greater than 0.02 mag. The other components remain constant over a
period of 10 hours to within about 0.05 mag. In the final 5 hours there is
some evidence (not conclusive) for variation of component D by about 0.1 mag.
Their results show that any fifth (central) component must be at least 6.5 mag
fainter than component A.  They also determined the astrometric properties of
the lens system; their values are systematically larger than
those of other investigators (by 0.1\% to 2.0\%), but there are reasons to
believe that the new results are more reliable.  The F336W filter was 
chosen for the observations because it corresponds to the redshifted
$\lya$ line of the quasar.  This filter might have allowed them to see
extended $\lya$ emission from the broad-line region (BLR) of the quasar
as $\lya$ arcs, and hence to determine the physical size of the BLR.
However, the quasar components in this filter are consistent with a point
source. They conclude that there cannot be any $\lya$ feature in the
image plane brighter than about 23.5 mag in F336W and further from the quasar
core than 100 mas. According to a lensing model by Rix, Schneider and Bahcall,
this would preclude any such features in the source plane further than 20 mas
($\sim 100 h^{-1}$ pc, assuming $q_0 = 0.5$) from the quasar core and brighter
than 25 mag before magnification.

Aperiodic optical variability is a common property of AGNs, though the
mechanism is still open to question.  To study the origin of the
optical-ultraviolet variability in AGNs, T. Kawaguchi, S. Mineshige (Kyoto),
M. Umemura (Tsukuba) and Turner compared light curves of two models to
observations of the quasar 0957+561.  In the starburst (SB) model, the random
superposition of supernovae in the nuclear starburst region produces aperiodic
luminosity variations, while in the disk-instability (DI) model, variability
is caused by instabilities in the accretion disk around a supermassive black
hole.  They calculated fluctuating light curves and structure functions,
$V(\tau)$, by Monte-Carlo simulations of the two models.  Each resultant
$V(\tau)$ possesses a power-law portion, $[V(\tau)]^{1/2} \propto
\tau^{\beta}$, at short time lags $\tau$.  The two models can be distinguished
by their logarithmic slope, $\beta$; $\beta \sim0.74$--0.90 in the SB model
and $\beta \sim0.41$--0.49 in the DI model, while the observed light curves
exhibit $\beta \sim 0.35$.  Therefore, the DI model is favored over the SB
model in this case.  In addition, they examined the time-asymmetry of the
light curves by calculating $V(\tau)$ separately for brightening and decaying
phases.  The two models exhibit opposite trends of time-asymmetry to some
extent, although the presently available light curve is not long enough to
test this prediction.

\subsection{Galaxies} 

\noindent
Over the past five years the evidence that massive black holes are commonly
found in the centers of inactive nearby galaxies has become extremely strong.
S. Tremaine, working with J. Magorrian (CITA), K. Gebhardt and D. Richstone
(Michigan), R. Bender (Munich), G. Bower, R. Green, T. Lauer (KPNO),
A. Dressler (Carnegie), S. Faber (Santa Cruz), C. Grillmair (JPL), and
J. Kormendy (Hawaii), has compiled the first large survey of the distribution
of black-hole masses, using a sample of 36 galaxies with HST photometry and
ground-based spectroscopy. They find that most galaxies contain central black
holes whose typical mass is $\sim0.6$\% of the mass of the spheroidal stellar
component of the galaxy; this result is entirely consistent with the
requirement that accretion onto these objects is the power source for quasars
and active galactic nuclei.

The same collaborative team has analyzed the structure of the centers of
early-type galaxies, using a sample of 61 elliptical galaxies and spiral
bulges with HST photometry. The photometric profiles are
combined with ground-based data on central velocity dispersions, total
luminosities, rotation velocities, and isophote shapes to explore correlations
among these parameters. Luminous galaxies ($M_V < -20.5$) show {\it core}
profiles, which have significant changes in their log-log surface-brightness
profiles at a break radius $r_b$.  Break radius and core luminosity are
approximately proportional to the analogous global parameters, effective
radius and total luminosity.  Cores follow a fundamental plane that parallels
the global fundamental plane but is 30\% thicker; some of this extra thickness
may be due to the effect of massive black holes on central velocity
dispersion.  Faint galaxies ($M_V > -22.0$) show steep, largely featureless
{\it power-law} profiles that lack cores. The centers of power-law galaxies
are up to 10$^{4-5}$ times denser in mass and luminosity than the cores of
large galaxies at a limiting radius of 10 pc. At intermediate magnitudes
($-22.0 < M_V < -20.5$), core and power-law galaxies coexist.  Central
properties correlate strongly with global rotation and shape: core galaxies
tend to be boxy and slowly rotating, whereas power-law galaxies tend to be
disky and rapidly rotating.  At intermediate magnitudes, the presence of a
core is a better predictor of boxiness and slow rotation than absolute
magnitude.

Ostriker and Tremaine investigated a longstanding paradox in galaxy dynamics:
why do galactic bars rotate with high pattern speeds, when dynamical friction
should rapidly couple the bar to the massive, slowly rotating dark halo? The
paradox may be resolved by considering the dynamical interactions between the
galactic disk and inhomogeneities in the dark halo. Plausible formation models
lead to a halo composed largely of tidal streamers. Dynamical friction between
these streamers and the disk spins up and flattens the inner halo, thereby
quenching the dynamical friction exerted by the halo on the bar. At the same
time the halo heats and thickens the disk, perhaps forming a rapidly rotating
bulge. More generally, gravitational scattering from tidal streamers or other
phase-wrapped inhomogeneities represents a novel relaxation process in stellar
systems, intermediate between violent relaxation and two-body relaxation,
which can isotropize the distribution function at radii where two-body
relaxation is not effective.

C. Murali (CITA) and Tremaine have examined the response of a galaxy to slowly
varying gravitational perturbations, such as those due to external tidal
fields. They have focused on the singular isothermal sphere, which is not only
a plausible model for a dark halo but also admits near-analytic solutions for
its linear response. For odd spherical harmonics, the response is identical to
the response of the analogous isothermal fluid system. For even spherical
harmonics, the response can be regarded as an infinite series of wavetrains in
$\log r$, implying alternating compression and rarefaction in equal
logarithmic radius intervals. Partly because of the oscillatory nature of the
solutions, tidal fields from external sources are not strongly amplified by an
intervening isothermal stellar system, except at radii $< 10^{-3.5}$ times
the satellite radius; at some radii the stellar system can even screen the
external tidal field in a manner analogous to Debye screening.
 
J. Touma (Texas) and Tremaine have devised a symplectic map to describe the
behavior of eccentric orbits in cuspy triaxial potentials, such as those
found near the centers of galaxies. The map correctly reproduces most of the
features of the surface of section for orbits in the non-axisymmetric
logarithmic potential, including the presence of minor resonant families found
by Schwarzschild (``fish'', ``bananas'', etc.), but is $\sim 10^3$ times
faster than orbit integration.

\subsection{Galactic astronomy and interstellar matter}

\noindent
Using OGLE data on the galactic bulge red clump stars, and the Hipparcos data
on the nearby red clump stars (over 1000 with parallax errors less than
10\%) Paczy\'nski and his former student K. Stanek (Harvard) determined the
distance to the Galactic Center to be $8.4\pm0.4\kpc$.  This method was
subsequently applied by various investigators to estimate distances to several
nearby galaxies.

Paczy\'nski found that the observed $V-I$ colors of red clump giants in the
galactic bulge are not correlated with the published determinations of
metallicity [Fe/H].  The reason for this puzzling result is not known, but low
accuracy of the [Fe/H] determinations is a possibility.  This problem has to
be solved before the red clump giants can be used as reliable standard
candles.

Draine has continued to work on the theoretical astrophysics of the
interstellar medium, with particular attention to problems connected with the
dynamics of interstellar dust grains, radiative transfer in X-ray irradiated
gas, and the physics of photodissociation fronts.

The process of grain alignment, whether by radiative torques or the
``classical'' Davis-Greenstein mechanism of paramagnetic dissipation,
characteristically involves ``crossover'' events, where the grain's angular
velocity component along the grain's principal axis of largest moment of
inertia changes sign.  These crossover events are critical to the process of
grain alignment, because the grain is easily disoriented during the period
when its angular momentum is small.  A. Lazarian and Draine have shown that
previous analyses of the crossover process overlooked the subtle but very
important effects of thermal fluctuations within the grain.  When these
fluctuations are taken into account, it is found that the larger grains---the
ones which are observed to be highly aligned in the interstellar medium---are
much less susceptible to disorientation during crossover than had previously
been believed.  As a result, it is found that even without radiative torques
acting, classical paramagnetic dissipation appears capable of achieving the
observed degree of grain alignment, for plausible assumptions regarding grain
properties.

Draine and Lazarian studied the rotational dynamics of dust grains and found
that very small grains could attain rotation rates as high as 100 GHz, and
that the small grain population would be expected to radiate electric dipole
emission with intensities and spectrum capable of accounting for the
``anomalous'' microwave emission which several groups have previously
discovered to be correlated with 100$\mu$m emission from interstellar dust.
They estimated the intensity of the rotational emission for various
interstellar environments and discussed the physical processes involved in the
rotational dynamics---these include damping by rotational electric dipole
radiation as well as by emission of infrared photons following heating by
absorption of starlight photons, and rotational excitation by collisions with
atoms and ions.

Draine and Lazarian have examined the physics of thermal fluctuations in
magnetic grains, and show that if the iron in interstellar grains resides in
ferrimagnetic or ferromagnetic materials, there can be strong magnetic dipole
radiation in the microwave region resulting from thermal fluctuations in the
magnetization of the magnetic materials. 

With W. Chiu (graduate student, Physics) Draine estimated the radiation
pressure resulting from trapped $\lya$ photons within a stellar
atmosphere subject to X-ray irradiation, as would be the case for a star
sufficiently close to an active galactic nucleus.  This required assessment of
the diffusion of resonance line photons, both in space and in frequency, to
relate the $\lya$ production rate to the $\lya$ pressure.  It is
found that for low gravity stars (e.g., supergiants) the radiation pressure
can approach $\sim$20\% of the gas pressure.  While a hydrostatic equilibrium
exists for a plane-parallel atmosphere, it seems likely that the radiation
pressure would have dramatic dynamical effects on a realistic stellar
atmosphere subject to unidirectional X-ray irradiation.

Draine pointed out that the dust-free gas clouds which have been proposed as a
``dark matter'' component of the Galaxy could be detectable through their
lensing effects on background stars.  If a significant fraction of the
gravitational mass of the Galaxy were contained in a halo population of
Jovian-mass cold clouds, they could be detected through their contribution to
lensing of stars in the Magellanic Clouds.  The light curves closely resemble
``gravitational lensing'' light curves in the high-magnification region, but
demagnification also is present (unlike gravitational lensing).  Accurate
photometry by existing programs to study microlensing could strongly constrain
the cloud population.  The hypothesized clouds could be definitively detected
during a gaseous lensing event by observation of far-red absorption lines of
H$_2$ in the stellar spectrum.

Lazarian worked with D. Pogosyan on the statistical description of galactic
atomic hydrogen, relating the statistics of 21~cm intensity to the statistics
of density and velocity. They showed that two distinctly different regimes of
interferometric observations of HI exist. The first one (``thick slicing'') is
sensitive to density fluctuations and the inversion suggested earlier by
Lazarian is applicable to it.  The other regime, which was called ``thin
slicing'', is sensitive to both velocity and density fluctuations and
therefore the velocity and density statistics can be disentangled. In another
paper Lazarian and Pogosyan studied how fluctuations of velocity deform
filaments that are observed in atomic hydrogen and how the observed
filamentary pattern depends on the spectrum of density and velocity.

Lazarian and P. Myers (Center for Astrophysics) studied the dynamics of
molecular clouds. This work was motivated by recent observations of cloud
contraction, and showed that dissipation of Alfven turbulence can account for
the observed velocities.

Lazarian and M. Efroimsky (Harvard) examined inelastic relaxation within
interstellar grains. This work was motivated by an earlier study by Lazarian
and Draine where it was found that internal relaxation can substantially alter
the efficiency of paramagnetic alignment. Lazarian and Efroimsky showed that
it is impossible to treat grain deformations arising from grain precession
using the formalism of acoustic modes and developed a new formalism that
accounts for the nearly isothermal character of grain deformations. They have
also modified their mathematical technique to account for adiabatic
deformations of asteroids and comets and calculated the rates of internal
dissipation for these bodies.

Jenkins and Sofia (Villanova) analyzed the 1048 and 1066\AA\ absorption
features from interstellar neutral argon in the ultraviolet spectra of nine
early-type stars observed by the Interstellar Medium Absorption Profile
Spectrograph (IMAPS) in 1993.  They found that for the stars $\zeta$~Pup,
$\gamma^2$~Vel and $\beta$~Cen, the abundance of argon to hydrogen is depleted
with respect to the solar or B-star abundance ratio by the logarithmic
reduction factors $D=-0.37\pm 0.09$ dex, $-0.18\pm 0.10$ dex and $-0.61\pm$
0.12 dex, respectively.  For the remaining stars, lower limits for $D$ were
obtained.  For the characteristically low-density lines of sight in this
study, it is unlikely that argon can be depleted onto dust grains.  Instead,
Jenkins and Sofia argued that the relatively large photoionization cross
section of neutral argon, compared to that of hydrogen, makes it much easier
to hide in its ionized form within regions that are partially ionized.  In
regions that are about half ionized, this effect can lower Ar~I/H~I by $-0.11$
to $-0.96$\,dex, depending on the energy of the photoionizing radiation and
its intensity divided by the local electron density.  They pointed out that
the observed values of Ar~I/H~I could be a good discriminant between energetic
photons and electron collisions as a source of ionization in H~I regions that
are partially ionized.

Jenkins measured the thermal pressure of the medium inside the Local Bubble [a
region that contains mostly hot ($\sim 10^6$K) gas out to a radius of $\sim
100$pc from the Sun] by analyzing absorption features of C~I in the spectrum
of $\delta$~Cyg ($l=79^o$, $b=+10^o$, $d=52$\,pc) recorded by the GHRS echelle
spectrograph on HST.  The fine-structure levels in the ground electronic state
of C~I can be excited by collisions, and from the population ratios of these
states Jenkins derived a thermal pressure range $10^{2.7}<p/k<10^{3.7}$, a
value that is lower than results reported by other investigators based on EUV
emission by the hot gas in front of cold clouds embedded in the Local Bubble.

HST Early Release Observations of the spectrum of HD72089 recorded at a
resolving power of $\lambda/\Delta\lambda=110,000$ for the Space Telescope
Imaging Spectrograph (STIS) investigation team were analyzed by Jenkins,
Tripp and E. Fitzpatrick.  This star is located behind gases that have been
accelerated by shocks within the Vela supernova remnant.  They identified
seven narrow components of C~I, some of which showed significant levels of
fine-structure excitation.  Absorption features from C~II, N~I, O~I, Si~II,
S~II and Ni~II appeared over a heliocentric velocity range from $-70$ to
$+130\,{\rm km~s}^{-1}$.  The analysis of the abundances of these species
indicates that some elements may be preferentially ionized to higher stages by
photoionizing radiation emitted by hot gas immediately behind the shock
fronts.  This effect may explain the remarkably low abundances of N~I and O~I.

Jenkins also collaborated with the STIS team in an interpretation of
interstellar features appearing in the O-type star CPD $-59^o2603$ in the
Carina Nebula, again taken from Early Release Observations.  This
investigation identified a heterogeneous collection of species, including CO
molecules, atoms in low and high stages of ionization, and features arising
from atoms at high velocity.  The investigation derived some properties of
accelerated gases within the nebula, as well as material residing in
foreground H~I and H~II regions.  The most conspicuous features were the Mg~I
and Mg~II profiles that showed many components over a heliocentric velocity
range $-235<v<+123\,{\rm km~s}^{-1}$.

L. Blitz (Berkeley), P. Teuben (Maryland), Dap Hartman and W. Butler Burton
(Leiden) and Spergel have suggested that the high--velocity clouds (HVCs) are
large clouds, with typical diameters of 25 kpc and containing $5\times10^7$
solar masses of neutral gas and $3\times10^8$ solar masses of dark matter,
falling onto the Local Group; altogether the HVCs contain 10$^{11}$ solar
masses of neutral gas. Their reexamination of the Local-Group hypothesis for
the HVCs connects their properties to the hierarchical structure formation
scenario and to the gas seen in absorption towards quasars.  They interpret
the more distant HVCs as dark matter ``mini--halos'' moving along filaments
towards the Local Group. Most poor galaxy groups should contain HI structures
to large distances bound to the group. The HVCs are local analogues of the
Lyman--limit clouds.  Their analysis of the HI data leads to the detection of
a vertical infall of low-velocity gas towards the plane. This implies that the
chemical evolution of the Galactic disk is governed by episodic infall of
metal-poor HVC gas that only slowly mix with the rest of the interstellar
medium.  The Local--Group infall hypothesis makes a number of testable
predictions: the HVCs should have sub-solar metallicities; their H$\alpha $
emission should be less than that seen from the Magellanic Stream; the clouds
should not be seen in absorption to nearby stars; and the clouds should be
detectable in both emission and absorption around other groups.

Knapp reviewed cold gas and star formation in elliptical galaxies at the
``Star Formation in Early-Type Galaxies'' conference in Guanajuato.  A
significant fraction of elliptical galaxies not in dense clusters is now
known to contain interstellar matter in all the familiar phases---hot, warm
and cold gas.  The cold gas content of elliptical galaxies is $<$ 1\% to about
10\% of the amounts found in spiral galaxies. In many elliptical galaxies,
cold gas is concentrated to the inner regions and has surface densities
comparable to, or greater than, the regions of highest star formation in the
Galactic disk.  The compact disks seen in the inner regions of some elliptical
galaxies may form from these clouds; however, the relative cold gas content
shows no statistical correspondence with galaxy morphology---core morphology,
boxiness/diskiness, or luminosity.  The cold and hot (X-ray emitting) gas
masses in elliptical galaxies are found to be anticorrelated.  The
observations show that in a statistical sense typical field elliptical
galaxies contain a small amount of cold interstellar matter: this is more
consistent with a long-lived, evolving interstellar medium than with recent
capture of the gas.

\subsection{Stellar systems}

\noindent
With S. S. Kim and H. M. Lee (Pusan), J. Goodman has examined the effect of a
mixture of stellar masses on the dynamical evolution of globular clusters
after core collapse.  Two components suffice to represent the stars most
common near the core: main-sequence stars at the turnoff and evolved
degenerates.  If the degenerates are individually more massive than the normal
stars and comprise a few percent or more of the total cluster mass, then they
dominate the core after core collapse and heat the cluster by three-body
processes.  Scaling laws for half-mass quantities are derived and confirmed by
Fokker-Planck calculations.  In agreement with speculations made by Goodman in
1993, the Fokker-Planck results show that two-component systems are stabler to
gravothermal oscillations than single-component models, and that the critical
parameter for gravothermal instability in postcollapse is the ratio of the
core and total cluster energies divided by the ratio of central to half-mass
relaxation times.  

Goodman and former Princeton graduate student E. Dickson have studied the
inexplicably rapid circularization of close main-sequence late-type binaries.
This is usually attributed to turbulent convective viscosity acting on the
equilibrium tide, but a previous paper by Goodman and Oh showed that the
convective mechanism is inadequate.  The new study examines the role of
tidally excited g-modes at the base of the convection zone, a mechanism often
invoked for early-type binaries but considered uninteresting for late types
because of the difficulty of damping the g modes in the radiative core.
Goodman and Dickson show that evolution of the resonant g-mode frequencies and
perhaps also nonlinear wave steepening near the center of the core can solve
the damping problem.  Even with efficient damping, however, the tidal coupling
is only strong enough to circularize binaries out to periods of about six days
on the main sequence, whereas observations indicate circularization out to at
least twelve days.

R. Nelson (Caltech) and Tremaine have applied linear response theory and the
fluc\-tu\-ation-dissipation theorem to stellar systems, to clarify the relation
between fluctuating gravitational forces and dynamical friction without using
traditional approximations such as local forces, instantaneous collisions,
neglect of self-gravity, small-angle deflections, etc. For isothermal stellar
systems, they derive an expression for the instantaneous dynamical friction
force, in terms of the correlation function of fluctuating forces, that does
not require {\it any} of these approximations.

\subsection{Stars}

\noindent
Knapp, K. Young (Center for Astrophysics), E. Lee (graduate student), and
A. Jorissen (Brussels) began a survey of thermal molecular line emission from
the envelopes of evolved mass-losing AGB stars to investigate a new
phenomenon: the presence of two winds with different expansion velocities.
CO(2-1) and CO(3-2) line emission was observed for 45 AGB stars at high
velocity resolution. Double winds are found in 20\% of the sample, and highly
asymmetric lines are found in six other stars.  The data tentatively suggest
that double winds occur when the star undergoes a change (pulsational mode,
chemical composition) and that the narrow components represent the onset of a
new phase of mass loss.

G. Wallerstein (Washington) and Knapp reviewed the subject of carbon stars.
The review discusses spectral classification, distances and luminosities
derived from the Hipparcos catalog, scale height, effective temperatures,
radii, molecular envelopes, mass-loss rates, abundances, and the properties of
special carbon stars such as the carbon dwarfs, R stars, and CH stars.

\u{Z}. Ivezi\' c and Knapp expanded the recent Jorissen and Knapp (1997) study
of S stars by including a detailed analysis of IRAS LRS data and variability
properties. The distribution of dust emission features in the 8-23 micron
region, and distribution of variability types (Miras vs.  SRb/Lb) both follow
the Jorissen and Knapp classification scheme based on five regions in
color-color diagrams. Although S stars show a greater variety of dust emission
features than O and C stars do, there is no firm evidence that S stars are
significantly different regarding their evolutionary status. The properties of
their infrared emission, mass-loss rates and outflow velocities are no more
different from those for O and C stars than what would be expected because of
somewhat different grain chemistry.  They also find that differences between S
type Miras and SRb/Lb variables are the same as the corresponding differences
for O and C stars.  Observed properties of O, C and S Miras closely agree with
those expected for a steady-state radiatively driven wind, while SRb/Lb
variables show an indication for a decrease of mass-loss rate during last
several hundred years.  It cannot be ruled out that the mass-loss rate is
changing periodically on comparable time scales, implying that the stars
oscillate between the Mira and SRb/Lb phases during their AGB evolution as
proposed by Kerschbaum et al. (1996). Such a possibility appears to be
supported by recent HST images of the Egg Nebula obtained by Sahai et
al. (1997), and the discovery of multiple CO winds by Knapp et al. (1998).

Ivezi\' c, Knapp, and M. Elitzur (Kentucky) have calculated detailed
self-consistent models of radiatively driven stellar outflows which couple the
radiative transfer and hydrodynamics equations. The circumstellar envelope,
which consists of gas and dust, is described as a two-component fluid to
account for relative drifts. Their results agree well with both molecular line
observations and infrared continuum spectra, thus providing strong evidence
that outflows around cool luminous late-type stars are radiatively driven.

\u{Z}. Ivezi\' c has continued to work in collaboration with Elitzur on the
scaling properties of the dust radiative transfer. Together with
A. Miroshnichenko (Pulkovo) and D. Vinkovi\' c (Kentucky), they apply
their results to the studies of medium- and high-mass young stellar objects
(e.g. Herbig Ae/Be stars, B[e] stars). In particular, they propose a simple
resolution of inconsistencies encountered when constraining the geometry of
dust around such stars. Spherical envelopes seem to be ruled out by submm
observations which imply visual optical depths of about 1000, yet the stars
are visible at optical wavelengths.  Similarly, an alternative proposal
invoking geometrically thin and optically thick disks cannot explain the
existence of silicate emission feature. Furthermore, neither model is capable
of explaining why for some sources the observed IR sizes decrease with
wavelength.  Ivezi\' c and collaborators show that an alternative model
involving both a spherical dusty envelope in free fall and an embedded
optically thick and geometrically thin disk can explain all available
observations. In particular, the puzzle of observed sizes decreasing with
wavelength is resolved as a consequence of the disk emission overtaking
envelope emission: the disk temperature decreases much faster with radius than
the envelope dust temperature, hence the effective size for disk emission is
smaller than for envelope emission.

\subsection{Planetary systems}

\noindent
One of the most remarkable features of the planets that have recently been
discovered around nearby stars is that they are found at small orbital
radii---in some cases less than 10\% of the Earth's---where gas giant planets
are not expected to form. Working with N. Murray and B. Hansen (CITA) and
M. Holman (Center for Astrophysics), Tremaine has argued that these planets
may have formed at much larger radii and then migrated inward to their current
orbits; the migration is caused by gravitational scattering of residual
planetesimals and appears to be inevitable in a system whose protoplanetary
disk is 10--100 times more massive than our own. Such massive disks are
plausible but not yet directly observed.

L. Malyshkin (graduate student) and Tremaine have examined the evolution of
highly eccentric, planet-crossing orbits in the restricted three-body problem
(Sun, planet, comet). Following such orbits for the lifetime of the solar
system using conventional integration techniques is a challenging task,
because of accumulated numerical errors and the large CPU time required. They
examined a simpler toy problem capturing most of the relevant physics: a
symplectic map in which the comet energy changes instantaneously at
perihelion, by an amount depending only on the azimuthal angle between the
planet and the comet at the time of perihelion passage. This approximate but
very fast mapping allowed them to explore the evolution of large ensembles of
long-period comets, and to compare their results on comet evolution with those
given by the diffusion approximation and by direct orbit integration of comet
orbits. They found that at long times the number of surviving comets is
determined by resonance sticking rather than a random walk; this result
greatly enhances the number of primordial Neptune-crossing planets that might
still be present in a scattered cometary disk. 

One of the most exciting results in planetary science in the 1990s was the
discovery of the Kuiper belt. By now over 60 Kuiper belt objects have been
discovered; these are $\sim 100\km$ sized bodies (comets? planetesimals?)
orbiting in a disk outside Neptune. Discovering these objects and tracking
them are two quite different tasks; without follow-up observations to recover
and track the objects over several seasons and determine their orbits, most
will be lost. Tremaine, B. Gladman (CITA), M. Holman (Center for
Astrophysics), W. Offut (Cloudcroft), B. Gillespie, C. Hastings, and K. Gloria
(APO) experimented with using the APO 3.5-meter telescope for this purpose;
they successfully recovered 4 Kuiper-belt objects with magnitude as faint as
$R=23.3\pm0.5$. 

\subsection{Magnetic Reconnection}

\noindent
Lazarian and Vishniac (Center for Astrophysics) studied magnetic
reconnection. They found an increase of the reconnection rates (as compared to
the Sweet-Parker rates) when magnetic fields are frozen into partially ionized
gas.  This increase is not sufficient to account for the magnetic diffusivity
required by contemporary dynamo theories, but is important for magnetic cloud
evolution in the interstellar medium. For instance, it can enable an efficient
detachment of the collapsed core from the large scale galactic magnetic
field. As a separate project Lazarian and Vishniac studied magnetic
reconnection in the presence of tearing modes. They found that the
reconnection velocity scales as $Rm^{-1/3}$, where $Rm$ is the magnetic
Reynolds number, while $Rm^{-1/2}$ is the prediction of Sweet-Parker theory.

R. Kulsrud and his students D. Uzdensky and T. Carter have continued to study
magnetic reconnection as a fast magnetohydrodynamic process, in collaboration
with an experimental study of the creation and merging of two plasma tori.
There are two established, and competing, theories for the process: the
Sweet-Parker theory and the Petschek theory.  In order to properly decide
between them it is necessary to study not only the reconnection layer itself,
but the separatrix layer into which the reconnected plasma must flow.  They
assume that the separatrix layer is infinitely conducting and under this
assumption they find that the flow in it can be treated independently of what
happens as the reconnected lines merge into the downstream equilibrium region.
For all possible initial conditions for the plasma in the layer they find that
the reconnection rate is invariably the Sweet-Parker rate.

The results of the simulations seem to contradict a generally accepted result
due to Cowling and Priest that the field lines at the very center of the
reconnection region must osculate rather than cross at a finite angle.  We
show that this is due to an unwarranted assumption about the analyticity of
the solution, and when this is corrected the results are consistent with the
numerical simulations.  Also this earlier Cowling-Priest result was extended
by Shivamoggi to include a small resistivity and again the result was inferred
to be osculating fields as the viscosity went to zero.  However, again the
conclusion was wrong since in this limit a boundary layer develops which
vitiates the conclusion.  The behavior of the boundary layer as viscosity goes
to zero is identical with numerical simulations carried out in the same limit.

The rather slow reconnection rate predicted by Sweet-Parker seems at variance
with the detailed X-ray pictures of solar flares observed with the Soho and
Yohcoh satellites.  It also is considerably slower than observed in the
laboratory experiment where the reconnection layer was measured to be thicker
than that predicted by their theory.  However, in both cases it can be shown
that conditions for normal resistivity are violated since for the predicted
current density of Sweet and Parker the critical current for excitation of
plasma waves is exceeded.  The plasma waves that one expects to be excited
produce anomalous resistivity that can be expected to explain most of the
discrepancy with the Sweet-Parker theory.

\subsection{Instrumentation and Software}

\noindent
The Princeton SDSS software group (R. Lupton, Strauss, Ivezi\' c, Fan, Gunn
and Knapp) have as their prime responsibility the photometric pipeline that
automatically reduces the data from the imaging camera on the SDSS 2.5-meter
telescope.  Observations with the imaging camera consist of continuous drift
scans, with the sky passing through five filters in turn.  These images are
cut into a series of frames and the data from a given area of the sky are
aligned; this is done by the ``serial stamp collecting'' pipeline, SSC.  The
set of five frames for a given field is analyzed by the Frames pipeline, which
carries out the most computationally intensive parts of the reduction, the
finding and measuring of objects.

Frames corrects the images for defects, cosmic rays and bad columns by
interpolation, subtracts the bias, flat-fields the data, finds objects,
measures them (position, brightness, size, shape, morphology), and classifies
them (assigning a set of likelihoods that the image shape fits a suite of
models---these likelihoods, plus priors, are the information on which
star-galaxy separation is based). Frames then ``deblends'' overlapping
objects, and writes out an object catalogue, a set of small cutouts (``atlas
images'') in each of the five bands for every detected object, the corrected
frames, masks for image defects and saturated pixels, and binned images with
the objects subtracted.

The positions (on the CCD) and the instrumental brightnesses are then
converted to position on the sky and magnitude using astrometric calibrations
from the astrometric array and photometric calibrations from a small telescope
(the ``Monitor Telescope'').  These calibration data are reduced through their
own software pipelines (``Astrom'', the responsibility of the U.S. Naval
Observatory, and ``MTpipe'', the responsibility of Fermilab).  Frames also
cuts atlas images at the positions of objects in catalogues at other
wavelengths (X-ray, radio, etc.) to aid in their optical identification.
Lupton is in charge of the Frames pipeline and responsible for most of the
code.

The Frames pipeline needs calibration data (point-spread function
characteristics, bias vector, flat field vectors, sky brightnesses) as input;
these need to be determined for the entire run and continuous from frame to
frame. The Postage Stamp Pipeline (PSP) determines these quantities for an
entire run, and interpolates them to the center of each frame.  As well as
cutting and aligning the frames, the SSC cuts subimages (``postage stamps'')
at the positions of bright stars found in the astrometric array (thereby
helping tie the astrometric, as well as photometric, solutions to the frames).
PSP analyzes these stamps; it selects stars and calculates the mean
point-spread function for a frame; this includes a calculation of the
scattering wings. The PSP also calculates the sky levels for an entire run.
Ivezi\'c is responsible for both SSC and PSP.

The photometric pipeline has been under development for about seven years,
first at Fermilab and later at Princeton.  It is required to reduce the data
at essentially the rate at which it is taken, 4.6 MB s$^{-1}$ (the current
performance is within a factor of two of this goal). To aid the development
process extensive simulations of the SDSS data have been carried out over the
years at Fermilab, and drift-scan data were taken during several nights in
September 1997 using the APO 3.5-meter telescope and the University of
Washington camera ``SPICam'' with a set of SDSS filters. 

Following first light for the SDSS camera in May 1998, test data taken with
the camera are now being reduced using this software, to test integration of
the different pipelines, and measure the astrometric and photometric accuracy.

The data are also being examined at Princeton by the above group, aided by
M. Vogeley, D. Schlegel, and a group of undergraduate summer students: J.
Pepper, J. Goldston, A. McDaniel, D. Freedman and K. Finlator. Sample tests
include: examining the PSF and its effect on image classification and
deblending; examining galaxy classification and the measured magnitudes for
galaxies; determining the sky brightness, limiting magnitude, camera/telescope
throughput and image quality, to investigate whether these meet expectation
and/or design; comparing SDSS data with those taken by other imaging
telescopes; investigating the distribution of stellar colors and brightnesses
to see if these are consistent with known data and Galaxy models and to
investigate selection criteria for spectroscopic observations of various types
of standard stars and of quasars; producing tools to monitor the quality of
the data flow and of the reductions.  This examination of the data by SDSS
scientists has already produced several science results: new clusters of
galaxies, new low surface brightness systems, and several new quasars. 

In addition, Lupton is helping develop and finish the software which operates
and monitors the SDSS instruments; Strauss is responsible for the independent
testing of the spectroscopic pipeline, which is being written by scientists at
the University of Chicago; Ivezi\' c and Goldston are working with Fermilab
astronomers on the photometric calibration software (MTpipe),
and on calibrating the recently acquired photometric data.

The SDSS consortium consists of the University of Chicago, the Fermi National
Accelerator Laboratory, the Institute for Advanced Study, the Japan
Participation Group (a group of scientists from the University of Tokyo), the
Johns Hopkins University, Princeton University, the U.S.  Naval Observatory
and the University of Washington.

Jenkins, Tripp, Wozniak, Sofia (Villanova) and Sonneborn (Goddard) have
completed most of the data analysis software for high-resolution UV spectra of
bright, early-type stars recorded by the Interstellar Medium Absorption
Profile Spectrograph (IMAPS) which flew on the ORFEUS-SPAS II mission in
November and December of 1996.  Major accomplishments in this effort included
the corrections for image distortions, derivations of wavelength scales,
determinations of background levels from various sources, and the creation of
routines to accomplish optimum extractions of the slightly overlapping echelle
orders.  The only remaining uncompleted task is the creation of routines that
can predict image shifts caused by changes in the Earth's magnetic field.
These shifts can be detected and compensated when the spectra of bright stars
are processed, but the prediction scheme is needed for stars that are too
faint to show features in individual frames.

Reale (electronics engineer) and Jenkins are developing circuitry to enable
the real-time detection of photoevents in image frames coming from the
windowless, electron-bombarded, intensified CCD image sensor that flew on
IMAPS.  This new capability will enable IMAPS to record spectra of faint stars
with a signal-to-noise ratio that is not degraded by the CCD read noise.  On
previous missions, the image information was accumulated by straight addition
of the analog signals from the video frames.  This mode is all right for
bright stars, but the spectra of fainter stars become lost in the CCD read
noise if the photoevents are not explicitly recognized before the signals are
combined.  The development of a photon detection capability for this far-UV
sensor should also be of potential benefit to experiments other than IMAPS.

As a member of the Space Telescope Imaging Spectrograph (STIS) team, Jenkins
has been involved in the effort to make STIS operate at its full potential.
In particular, he has evaluated the prospects of artificially suppressing the
large dark count rate for the near-UV MAMA detector by temperature cycling.
This scheme may be implemented after a cooler is installed during the third
servicing mission for HST, thus enabling this detector to record better
quality spectra of faint sources in the near-UV band.

Flux estimates for faint sources or transients are systematically biased high
because there are far more truly faint sources than bright.  D. W. Hogg
(Institute for Advanced Study) and Turner have developed a maximum-likelihood
method to correct for this effect as a function of signal-to-noise ratio and
the (true) slope of the number-flux relation. The implications of these
corrections for analyses of surveys are substantial; the most important is
that sources identified at signal-to-noise ratios of four or less have
essentially unknown fluxes.

C. Alard (Paris Observatory) and R. Lupton have developed new photometric
software to detect variable sources, based on the ``image subtraction''
concept, which offers substantial advantages over conventional methods based
on source fitting. This software will be implemented in the automated OGLE
data pipeline.

Spergel belongs to a collaboration led by J. Mather that is exploring the
possibility of developing a Far-Infrared Space Interferometer.  In light of
recent technical advances it is now practical to consider the possibility of
placing a far-infrared interferometer in orbit early in the next century. Such
a mission would provide unprecedented access to the far-IR (50--300 microns)
at high spatial resolution, enabling studies of primordial gas clouds
collapsing to form the first galaxies, the chemical evolution of the universe,
and detailed observations of dust-enshrouded Galactic protostars, star-forming
regions in nearby galaxies, and AGNs.

\section{References}

\begin{enumerate}

\item {\bf Bahcall, N.A.}, 1997, in Critical Dialogues in Cosmology,
ed. N. Turok (Singapore: World Scientific), 221

\item {\bf Bahcall, N.A.}, 1998, in New Horizons from Multi-Wavelength Sky
Surveys, IAU Symposium 179, ed. B. J. McLean, D. A. Golombek, J.J.E. Hayes,
H. E. Payne (Dordrecht: Kluwer), 317

\item {\bf Bahcall, N.A.}, 1998, in Large Scale Structure: Tracks and Traces,
Proceedings of the 12th Potsdam Cosmology Workshop, ed. V. M\"uller,
S. Gottl\"ober, J. P. M\"ucket \& J. Wambsganss (Singapore: World Scientific)

\item {\bf Bahcall, N.A.}, 1998, in proceedings of 14$^{\rm th}$ IAP
Colloquium, Wide Field Surveys in Cosmology, ed. Y. Mellier \& S. Colombi

\item {\bf Bahcall, N.A.}, 1998, in proceedings of MPA/ESO Cosmology
Conference, Evolution of Large Scale Structure, ed. S. White (Vledder: Twin
Press)  

\item {\bf Bahcall, N.A.} \& {\bf Fan, X.}, 1998, in Proc. Nat. Acad. 
Sci., 95, 5956

\item {\bf Bryan, G.}, \& Norman, M. L. 1997, in 12$^{\rm th}$ Kingston
Meeting on Theoretical Astrophysics: Computational Astrophysics,
ed. D. A. Clarke \& M. J. West (San Francisco: ASP), 363

\item {\bf Bryan, G. L.}, \& Norman, M. L. 1997, in proceedings of workshop on
Structured Adaptive Mesh Refinement Grid Methods, ed. N. Chrisochoides (IMA
workshop series)

\item {\bf Bryan, G. L.}, \& Norman, M. L. 1998, ApJ, 495, 80

\item {\bf Cen, R.}, 1998, ApJ, 498, L99.

\item {\bf Cen, R.}, Phelps, S., Miralda-Escud{\'e}, J., \& {\bf Ostriker,
J. P.} 1998, ApJ, 496, L577

\item Chiu, W. A., {\bf Ostriker, J. P.}, \& {\bf Strauss, M. A.} 1998, ApJ,
494, 479 

\item {\bf Colley, W.} 1997, ApJ, 489, 471

\item {\bf Colley, W.}, {\bf Gnedin, O.}, {\bf Ostriker, J. P.}, \& Rhoads,
J. 1997, ApJ, 488, 579 

\item Cornish, N. J., {\bf Spergel, D. N.}, \& Starkman, G. 1998,
Phys. Rev. D, 57, 5982

\item Cornish, N. J., {\bf Spergel, D. N.}, \& Starkman, G. 1998,
 Proc. Nat. Acad. Sci., 95, 82

\item Dalcanton, J. J., {\bf Spergel, D. N.}, {\bf Gunn, J. E.}, Schmidt, M.,
\& Schneider, D. P. 1997, AJ, 114, 635

\item {\bf Draine, B. T.} \& {\bf Lazarian, A.} 1998, ApJ, 494, L19

\item Faber, S. M., {\bf Tremaine, S.}, Ajhar, E. A., Byun, Y.-I., Dressler,
A., Gebhardt, K., Grillmair, C., Kormendy, J., Lauer, T. R., \& Richstone,
D. 1997, AJ 114, 1171

\item {\bf Fan, X.}, {\bf Bahcall, N.A.}, \& {\bf Cen, R.}, 1997, ApJ, 490,
 L123

\item {\bf Gnedin, O. Y.} 1997, ApJ, 487, 663

\item {\bf Gnedin, O. Y.} 1997, ApJ, 491, 69

\item {\bf Goldberg, D. M.} 1998, ApJ, 498, 156.

\item {\bf Goldberg, D. M.}, \& {\bf Strauss, M. A.} 1998, ApJ, 495, 29

\item {\bf Goldberg, D. M.}, \& {\bf Wozniak, P.} 1998, AcA, 48, 19

\item {\bf Goodman, J.} 1997, New Astronomy, 2, 449

\item {\bf Goodman, J.}, \& {\bf Oh, S.P.} 1997, ApJ, 486, 403

\item Governato, F., Moore, B., {\bf Cen, R.}, Stadel, J., Lake, G., \& Quinn,
T. 1997, New Astronomy, 2, 91

\item  Guzzo, L., {\bf Strauss, M. A.}, Fisher, K. B.,
Giovanelli, R., \& Haynes, M. 1997, ApJ, 489, 37

\item Hogg, D. W. \& {\bf Turner, E. L.} 1998, PASP, 110, 727

\item {\bf Ivezi\' c, \v Z.}, \& Christodoulou, D. 1997, ApJ, 486, 818

\item {\bf Ivezi\' c, \v Z.}, \& Elitzur, M. 1997, MNRAS, 287, 799

\item {\bf Ivezi\' c, \v Z.}, Groenewegen, M.A.T., Men'shchikov, A., \&
Szczerba, R. 1997, MNRAS, 291, 121

\item {\bf Jenkins, E. B.} 1998, in The Local Bubble and Beyond,
ed. D. Breitschwerdt, M. J. Freyberg, \& J. Tr\"umper, IAU Colloquium 166
(Dordrecht: Kluwer), 33

\item {\bf Jenkins, E. B.}, {\bf Tripp, T. M.}, Fitzpatrick, E. L., Lindler,
D., Danks, A. C., Beck, T. L., Bowers, C. W., Joseph, C. L., Kaiser, M. E.,
Kimble, R. A., Kraemer, S. B., Robinson, R. D., Timothy, J. G., Valenti,
J. A., \& Woodgate, B. E. 1998, ApJ, 492, L147

\item Ji, H., Yamada, M., Hsu, S., \& {\bf Kulsrud, R.} 1998,
Phys. Rev. Lett., 80, 3256

\item Jorissen, A., \& {\bf Knapp, G. R.} 1998, A\&AS, 129, 363

\item {\bf Kepner, J. V.}, Babul, A., \& {\bf Spergel, D. N.} 1997, ApJ, 487,
61

\item {\bf Kim, R.S.J.}, \& {\bf Strauss, M. A.} 1998, ApJ, 493,
39

\item Kim, S. S., Lee, H.-M., \& {\bf Goodman, J.} 1998, ApJ, 495, 786

\item Kimble, R. A.,$\ldots$, {\bf Jenkins, E. B.}, et al. 1998, ApJ, 492, L83

\item {\bf Knapp, G. R.} 1997, Sky and Telescope, 94, 40

\item {\bf Knapp, G.R.} 1998, Zenit, 9, 387

\item {\bf Knapp, G.R.}, {\bf Gunn, J.E.}, Margon, B., {\bf Strauss, M.A.},
{\bf Lupton, R.H.}, Szokoly, G., \& Szalay, A.S. 1997, The Sloan Digital Sky
Survey CD (Astrophysical Research Consortium)

\item {\bf Knapp, G. R.}, Jorissen, A., \& Young, K. 1997, A\&A, 326, 318

\item {\bf Knapp, G.R.}, Young, K., {\bf Lee, E.}, \& Jorissen, A. 1998, ApJS,
117, 209

\item {\bf Kulsrud, R.} 1997, in Critical Dialogues In Cosmology,
ed. N. Turok (Singapore: World Scientific), 328

\item {\bf Kulsrud, R.} 1997, Physics of Plasmas, 4, 3960

\item {\bf Kulsrud, R.} 1997, Physics of Plasmas, 5, 1599

\item Kundi\'c, T., {\bf Turner, E. L.}, {\bf Colley, W.}, {\bf Gott, J. R.},
Rhoads, J., {\bf Wang, Y.}, Bergeron, L. E., Gloria, K. A., Long, D. C.,
Malhotra, S., \& Wambsganss, J. 1997, ApJ, 482, 75

\item {\bf Lazarian, A.} 1998, MNRAS, 293, 208

\item {\bf Lazarian, A.} 1998, in Statistical Properties of HI,
ed. J. Franco \& A. Carraminana (Cambridge: Cambridge University Press), 104

\item {\bf Lazarian, A.}, \& {\bf Draine, B. T.} 1997, ApJ, 487, 248

\item {\bf Lazarian, A.}, Goodman, A., \& Myers, P. 1997, ApJ, 490, 273

\item {\bf Lazarian, A.}, \& Pogosyan, D. 1997, ApJ, 491, 200

\item {\bf Lazarian, A.}, \& Roberge, W. 1997, MNRAS, 287, 941

\item {\bf Li, L. X.}, \& {\bf Gott, J. R.} 1998, Phys. Rev. Lett., 80, 2980

\item Magorrian, J., {\bf Tremaine, S.}, Richstone, D., Bender, R., Bower, G.,
Dressler, A., Faber, S. M., Gebhardt, K., Green, R., Grillmair, C., Kormendy,
J., \& Lauer, T. 1998, AJ 115, 2285

\item Malhotra, S., Rhoads, J. E., \& {\bf Turner, E. L.} 1998, in 
Proceedings of the Eighteenth Texas Symposium on Relativistic Astrophysics and
Cosmology, ed. A. Olinto, J. Frieman, \& D. Schramm (Singapore: World
Scientific), 520

\item Murali, C., \& {\bf Tremaine, S.} 1998, MNRAS 296, 749

\item Murray, N., Hansen, B., Holman, M. \& {\bf Tremaine, S.} 1998, Science
279, 69

\item {\bf Ostriker, J. P.}, \& {\bf Gnedin, O. Y.} 1997, ApJ, 487, 667

\item {\bf Paczy\'nski, B.} 1998, ApJ, 494, L23

\item {\bf Paczy\'nski, B.} 1998, ApJ, 494, L45

\item {\bf Paczy\'nski, B.} 1998, in 4th Huntsville GRB Symposium, AIP
Conf. Proc. 428, ed. C. A. Meegan, R. D. Preece, \& T. H. Koshut (New York:
American Institute of Physics), 783

\item {\bf Paczy\'nski, B.} 1998, AcA, 48, 405

\item {\bf Paczy\'nski, B.}, \& Kouveliotou, C. 1997, Nature, 389, 548

\item {\bf Paczy\'nski, B.}, \& Stanek, K. Z. 1998, ApJ,494, L219

\item Park, M.-G., \& {\bf Gott, J.R.} 1997, ApJ, 489, 476

\item Peebles, P.J.E., Schramm, D. N., {\bf Turner, E. L.}, \& Kron, R. G.
1998, in Scientific American Presents the Magnificent Cosmos (San Francisco:
Freeman), 86.

\item Pietrzynski, G., Udalski, A., Kubiak, M., Szymanski, M., {\bf Wozniak,
P.}, \& Zebrun, K. 1998, AcA, 48, 175

\item Rauch, M., Miralda-Escud{\'e}, J., Sargent, W.L.W., Barlow, T. A.,
Weinberg, D. H., Hernquist, L., Katz, R., {\bf Cen, R.}, \& {\bf Ostriker,
J. P.} 1997, ApJ, 489, 7

\item Savage, B. D., {\bf Tripp, T. M.}, \& Lu, L. 1998, AJ, 115, 436

\item Sembach, K. R., Savage, B. D., \& {\bf Tripp, T. M.} 1997, ApJ, 480, 216

\item Sigad, Y., Eldar, A., Dekel, A., {\bf Strauss, M. A.},
\& Yahil, A. 1998, ApJ, 495, 516

\item Sofia, U. J., \& {\bf Jenkins, E. B.} 1998, Ap.J, 499, 951

\item {\bf Spergel, D. N.} 1998, Class. Quantum Grav., 15, 1.

\item {\bf Spergel, D. N.}, \& Pen, U. 1998, ApJ, 491, L67

\item {\bf Spergel, D. N.}, \& Zaldarriaga, M. 1997, Phys. Rev. Lett., 79,
 2180

\item {\bf Strauss, M. A.}, {\bf Ostriker, J. P.}, \& {\bf Cen, R.} 1998,
ApJ, 494, 20

\item Suginohara, M., Suginohara, T., \& {\bf Spergel, D. N.} 1998, ApJ, 495,
511

\item Tegmark, M., Hamilton, A.J.S., {\bf Strauss, M. A.}, {\bf Vogeley,
M. S.}, \& Szalay, A. S. 1998, ApJ, 499, 555

\item Touma, J., \& {\bf Tremaine, S.} 1997, MNRAS 292, 905

\item {\bf Tripp, T. M.}, Lu, L., \& Savage, B. D. 1997, ApJS, 112, 1

\item {\bf Tripp, T. M.}, Lu, L., \& Savage, B. D. 1997, in Structure \&
Evolution of the Intergalactic Medium from QSO Absorption Line Systems,
ed. P. Petitjean \& S. Charlot (Paris: Editions Fronti\`eres), 452

\item {\bf Tripp, T.M.}, Lu, L., \& Savage, B.D. 1998, in The Scientific
Impact of the Goddard High Resolution Spectrograph, ed. J. C. Brandt,
T. B. Ake, \& C. C. Petersen (San Francisco: ASP), 261

\item {\bf Turner, E. L.} \& Umemura, M. 1997, ApJ, 483, 603

\item Udalski, A., Szymanski, M., Kubiak, M., Pietrzynski, G., {\bf Wozniak,
P.}, \& Zebrun, K. 1998, AcA, 48, 147

\item Udalski, A., Szymanski, M., Kubiak, M., Pietrzynski, G., {\bf Wozniak,
P.}, \& Zebrun, K. 1998, AcA, 48, 1

\item Udalski, A., Szymanski, M., Kubiak, M., Pietrzynski, G., {\bf Wozniak,
P.}, \& Zebrun, K. 1997, AcA, 47, 431

\item {\bf Ulmer, A.}, {\bf Paczy\'nski, B.}, \& {\bf Goodman, J.} 1998, A\&A,
333, 379

\item {\bf Vogeley, M. S.}, 1997, in The Evolving Universe: Selected Topics
on Large-Scale Structure and on the Properties of Galaxies, ed. D. Hamilton
(Dordrecht: Kluwer), 395

\item Walborn, N. R., Danks, A. C., Sembach, K. R., Bohlin, R. C., {\bf
Jenkins, E. B.,} Gull, T. R., Lindler, D. J., Feggans, J. K., Hulbert, S. J.,
Linsky, J., Hutchings, J. B., \& Joseph, C. L. 1998, ApJ, 492, L169

\item Wambsganss, J., {\bf Cen, R.}, \& {\bf Ostriker, J. P.} 1998, ApJ, 494,
29

\item {\bf Wang, Y.}, {\bf Spergel, D. N.}, \& {\bf Turner, E.L.} 1998, ApJ,
498, 1

\item {\bf Wang, Y.},  \& {\bf Turner, E. L.} 1997, Phys. Rev. D, 56, 724

\item {\bf Wang, Y.}, \& {\bf Turner, E. L.} 1997, MNRAS, 292, 863

\item {\bf Wang, Y.}, \& {\bf Turner, E. L.} 1998, in Proceedings of the
Eighteenth Texas Symposium on Relativistic Astrophysics and Cosmology,
ed. A. Olinto, J. Frieman, \& D. Schramm (Singapore: World Scientific),
517

\item {\bf Wozniak, P.}, \& {\bf Paczy\'nski, B.} 1997, ApJ, 487, 55

\item Yamada, M., Ji, H., Hsu, S., Carter, T., {\bf Kulsrud, R.}, Bretz, N.,
Jobes, F., Ono, Y., \& Perkins, F. 1997, Physics of Plasmas, 4,
1936

\item Zaldarriaga, M., {\bf Spergel, D. N.}, \& Seljak, U. 1997, ApJ, 488, 1

\end{enumerate}

\end{document}